\documentclass[12pt]{article}
\usepackage{epsfig}
\def\beq{\begin{equation}}
\def\eeq{\end{equation}}
\def\beqn{\begin{eqnarray}}
\def\eeqn{\end{eqnarray}}
\def\as{\alpha_{\mbox{\tiny S}}}
\def\ee{e^+e^-}
\def\bom#1{{\mbox{\boldmath $#1$}}}
\def\Ejet{E_{\mbox{\scriptsize jet}}}
\def\eps{\epsilon}
\def\gtap{\raisebox{-.45ex}{\rlap{$\sim$}} \raisebox{.45ex}{$>$}}
\def\half{{\textstyle {1\over2}}}
\def\nlf{n_f}
\def\tdt{s\frac{\partial}{\partial s}}
\def\thrhf{{\textstyle {3\over2}}}
\def\VEV#1{\left\langle #1\right\rangle}
\def\ycut{y_{\mbox{\scriptsize cut}}}

\begin{document}
%\begin{titlepage}
\begin{flushright}
CERN--TH/99--408\\
Cavendish-HEP-99/18\\
hep-ph/9912399
\end{flushright}              
\vspace*{\fill}
\begin{center}
{\Large \bf Theoretical Aspects of Particle Production
\footnote{Lectures at International Summer School on Particle Production
Spanning MeV and TeV Energies, Nijmegen, The Netherlands, August 1999.}}
\end{center}
\par \vskip 5mm
\begin{center}
 B.R. Webber\\
 Theory Division, CERN, 1211 Geneva 23, Switzerland\\
 and\\
 Cavendish Laboratory, University of Cambridge,\\
 Cambridge CB3 0HE, U.K.\footnote{Permanent address.}
\end{center}
\par \vskip 2mm

\section{Introduction}
In these lectures I shall describe some of the latest data on
particle production in high-energy collisions and compare them
with theoretical calculations and models based on QCD.
The discussion will concentrate mainly on hadron distributions
in jets, which are the manifestation at the hadronic level of
hard (high-momentum-transfer) scattering of the partons
(quarks and gluons) which are the fundamental fields of QCD.

In sect.~\ref{sec:frag}, the connection between parton
and hadron distributions is made more precise using the
concept of fragmentation functions. I concentrate in particular
on the region of small momentum fractions, where interesting
characteristic features of QCD are manifest. Next, in
sect.~\ref{sec:hadro}, the various available models
for the conversion of partons into hadrons are reviewed.
In sect.~\ref{sec:yields}, the predictions of theory and
models are compared with experimental data. After that,
sect.~\ref{sec:qg} focuses on new
data that show clearly the differences between jets that
originate from quark and gluon fragmentation.

Deep inelastic lepton scattering (DIS) at HERA is a copious source of jets;
sect.~\ref{sec:DIS} discusses new results on the properties of jets
in the so-called current and target fragmentation regions.

In sect.~\ref{sec:heavy} I discuss new data on the
fragmentation of heavy ($b$) quark jets into B mesons, and
finally sect.~\ref{sec:conc} draws some brief conclusions.

Many of the topics mentioned here are discussed more fully in
ref.~\cite{Ellis:1991qj}. In order to bring the discussion 
up to date, I have tried wherever possible to refer to
the very latest experimental data. Therefore many of
the references and figures concern preliminary data shown only
at conferences, in particular at the International Europhysics
Conference on High Energy Physics (EPS-HEP 99) held in Tampere,
Finland, in July 1999. The cited contributed papers can be
found through the EPS-HEP 99 web page \cite{Tampere} or, in the
case of the large collaborations, through the collaboration pages
\cite{ALEPH}-\cite{D0}.

\section{Jet fragmentation -- theory}\label{sec:frag}
\begin{figure}\begin{center}
\epsfig{file=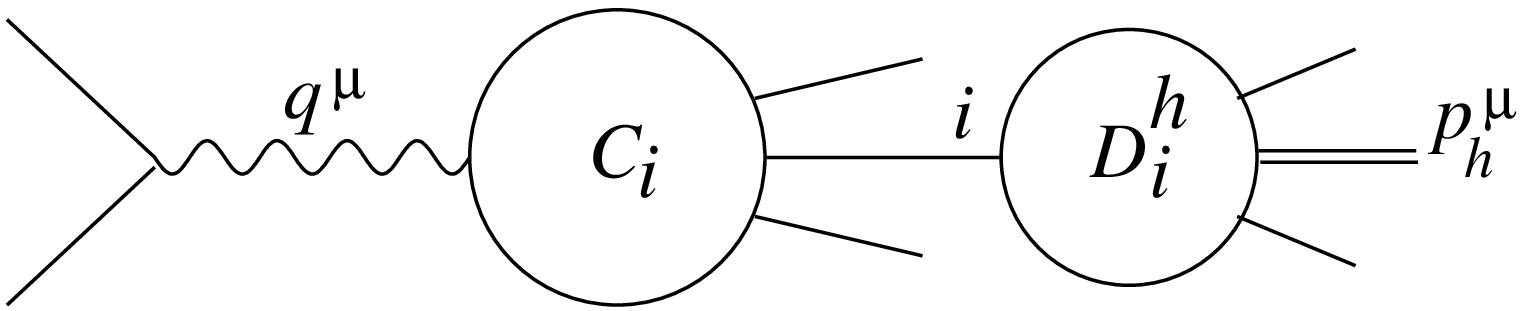,width=8cm}
\caption{Factorization structure of $\ee\to hX$.}
\label{fig:factn}\end{center}\end{figure}

We start with the basic factorization structure \cite{Collins:1987pm}
of the single-particle inclusive distribution, e.g.\ in $\ee\to hX$
(fig.~\ref{fig:factn}):
\beq\label{Fhxs}
F^h(x,s) = \sum_i\int_x^1\frac{dz}{z}C_i(z,\as(s))D^h_i(x/z,s)\;,
\eeq
\beq
s=q^2\;,\qquad x=2p_h\cdot q/q^2 = 2E_h/E_{cm}
\eeq
where $C_i$ are the {\em coefficient functions}
for this particular process (including all selection cuts etc.)
and $D^h_i$ is the universal {\em fragmentation function} \cite{Jakob}
for parton $i\to$ hadron $h$.

Although universal, fragmentation functions are factorization scheme
dependent \cite{Collins:1987pm}.  If one tries to calculate them in
perturbation theory, one encounters divergences associated with the
propagation of partons over long distances. In reality, however,
partons are confined and cannot travel long distances.  The
perturbative divergences can be collected into overall factors that
are replaced by non-perturbative factors taken from experiment. In
this way the incorrect long-distance behaviour of perturbation theory
is replaced by the correct long-distance features of QCD. However,
the factorization of the divergent terms is ambiguous: one can choose
to include different finite parts as well.  This is the
{\em factorization scheme ambiguity}.
To specify the scheme requires calculation of the coefficient functions to
(at least) next-to-leading order.  This has only been done in a few cases.
Thus there is need for theoretical work to make full use of the data on
fragmentation functions.

In certain kinematic regions, higher-order corrections are
enhanced by large logarithms, which need to be {\em resummed} to all
orders. Large logarithms of ratios of invariants may appear inside
the coefficient functions $C_i$, for example in multi-jet processes when
the angles between jets become small. In some cases these can be absorbed
into a change of scale in the fragmentation functions.
Examples of this will be encountered in
sects.~\ref{sec:yields} and \ref{sec:qg}.

The fragmentation functions $D_i^h$ are not perturbatively calculable but
their $s$-dependence (scaling violation) is given by the DGLAP evolution
equation \cite{Dokshitzer:1977sg,Gribov:1972rt,Altarelli:1977zs}:
\beq\label{APFi}
s\frac{\partial}{\partial s} D^h_i(x,s)
= \sum_j\int_x^1\frac{dz}{z}P_{ji}(z,\as(s))D^h_j(x/z,s)
\eeq
where $P_{ji}$ is the parton $i\to j$ {\em splitting function}. 
Thus fragmentation functions can be parametrized at some fixed scale
$s_0$ and then predicted at other energies \cite{Jakob}.
%The splitting functions $P_{ji}$ are scheme dependent in higher orders. 

The most common strategy for solving the DGLAP
equation is to take moments (Mellin transforms) with respect to $x$:
\beq\label{Mellin}
\tilde D(N,s) = \int_0^1 dx\; x^{N-1}\; D(x,s)\; ,
\eeq
the inverse Mellin transform being
\beq\label{invMell}
D(x,s) = \frac{1}{2\pi i}\int_C dN\; x^{-N}\; \tilde D(N,s)\;,
\eeq
where the integration contour $C$ in the complex $N$ plane
is parallel to the imaginary axis and to the right of all
singularities of the integrand.  After Mellin transformation,
the convolution on the right-hand side of eq.~(\ref{APFi})
becomes simply a product.

The moments $\tilde P_{ji}$ of the splitting functions 
are called {\em anomalous dimensions}, usually denoted by
$\gamma_{ji}(N,\as)$. They have perturbative expansions of the form
\beq
\gamma_{ji}(N,\as) = \sum_{n=0}^\infty
\gamma^{(n)}_{ji} (N)\left(\frac{\as}{2\pi}\right)^{n+1}\;.
\eeq

We can consider fragmentation function
combinations which are non-singlet in flavour, such as
$D_V=D_{q_i} - D_{\bar q_i}$ or $D_{q_i} -D_{q_j}$. In these
combinations the mixing with the flavour-singlet gluon drops
out and for a fixed value of $\as$ the solution is simply
\beq
\tilde D_V(N,s) = \tilde D_V(N,s_0)
\left(\frac{s}{s_0}\right)^{\gamma_{qq}(N,\as)}  \;.
\eeq

For a running coupling $\as(s)$, the scaling violation is no
longer power-behaved in $s$. The lowest-order form of the running
coupling is
\beq\label{s14e7}
\as(s) = \frac{1} {b \ln ( s /\Lambda^2)}
\eeq
with $b=(11 C_A-2\nlf )/12\pi$, where $C_A=3$ for QCD and
$\nlf$ is the number of `active' quark flavours, i.e.\ the
number with $m_q^2\ll s$.  Using this we find the solution
\beq \label{scaviol}
\tilde D_V(N,s) = \tilde D_V(N,s_0) \left(\frac{\as
(s_0)}{\as (s)}\right)^{d_{qq}(N)} 	, \;\; d_{qq}(N)=
\frac{\gamma^{(0)}_{qq} (N)}{2\pi b}\; ,
\eeq
which varies like a power of $\ln s$.

For the singlet fragmentation function
\beq
D_S=\sum_i (D_{q_i} + D_{\bar q_i})
\eeq
we have mixing with the fragmentation of the gluon and the
evolution equation becomes a matrix relation of the form
\beq\label{singev}
\tdt \left(\begin{array}{c} D_S \\ D_g \end{array}\right)
= \left(\begin{array}{cc} \gamma_{qq} & 2\nlf\gamma_{gq} \\
\gamma_{qg} & \gamma_{gg} \end{array}\right)
\left(\begin{array}{c} D_S \\ D_g \end{array}\right)\; .
\eeq
The anomalous dimension matrix in this equation has two
real eigenvalues $\gamma_\pm$ given by
\beq
\gamma_{\pm}=\half[\gamma_{gg}+\gamma_{qq}\pm
\sqrt{(\gamma_{gg}-\gamma_{qq})^2+8\nlf\gamma_{gq}\gamma_{qg}}]\;. 
\eeq
Expressing $D_S$ and $D_g$ as linear combinations of the
corresponding eigenvectors $D_+$ and $D_-$, we find that they
evolve as superpositions of terms of the form (\ref{scaviol})
with $\gamma_+$ and $\gamma_-$ in the place of $\gamma_{qq}$.

At small $x$, corresponding to $N\to 1$, the $q\to g$ and $g\to g$
anomalous dimensions have a singularity,
\beq\label{gammas}
\gamma_{gq}\to \frac{C_F\as}{\pi(N-1)} + {\cal O}(\as^2)\;,\qquad
\gamma_{gg}\to \frac{C_A\as}{\pi(N-1)} + {\cal O}(\as^2)
\eeq
($C_F=4/3$), and we find
$\gamma_+\to \gamma_{gg}\to\infty$, $\gamma_-\to\gamma_{qq}\to 0$.
Thus the low-$x$ region requires special treatment, as we discuss in the
following subsection.

\subsection{Small-x fragmentation}
At small $x$, multiple soft parton emission gives rise to terms enhanced by
up to two powers of  $\ln x$ for each power of $\as$. The leading
enhanced terms can be resummed by changing the DGLAP equation (\ref{APFi}) to
\beq\label{APtlt}
\tdt D^h_i(x,s)
= \sum_j\int_x^1\frac{dz}{z}P_{ji}(z,\as(s))D^h_j(x/z,z^2s)\;.
\eeq
The fact that the scale on the right-hand side should be $z^2s$ rather
than $s$ follows from {\em angular ordering} of successive
parton emissions \cite{Ermolaev:1981cm,Mueller:1981ex}.

For simplicity, consider first the solution of eq.~(\ref{APtlt})
for gluon fragmentation, taking $\as$ fixed and neglecting the sum
over different partons.  Then taking moments as before we
have
\beq\label{APtltN}
\tdt\tilde D(N,s) = \int_0^1 dz\,z^{N-1} P(z,\as)
\tilde D(N,z^2 s)\;.
\eeq
Now if we try a solution of the form
\beq\label{DNtlt}
D(N,s) \propto s^{\gamma(N,\as)}
\eeq
we find that the anomalous dimension $\gamma(N,\as)$ must satisfy
the implicit equation
\beq\label{gamN}
\gamma(N,\as) = \int_0^1 dz\,z^{N-1+2\gamma(N,\as)} P(z,\as)\; .
\eeq

When $N-1$ is not small, we can neglect the $2\gamma(N,\as)$ in
the exponent of eq.~(\ref{gamN}) and then we obtain the usual
explicit formula for the anomalous dimension.  For $N\simeq 1$,
the region we are interested in, the $z\to 0$ behaviour
$P_{gg} \to C_A\as/\pi z$ dominates, which implies that near $N=1$
\beq
\gamma_{gg}(N,\as) = \frac{C_A\as}{\pi}
\frac{1}{N-1+2\gamma_{gg}(N,\as)}
\eeq
and hence
\beqn\label{gamggN}
& &\gamma_{gg}(N,\as) \;=\;\frac{1}{4}\left[\sqrt{(N-1)^2 +
\frac{8C_A\as}{\pi}} - (N-1)\right]\nonumber \\
& & = \sqrt{\frac{C_A\as}{2\pi}} -\frac{1}{4}(N-1)
+\frac{1}{32}\sqrt{\frac{2\pi}{C_A\as}}(N-1)^2 + \cdots
\eeqn

Thus for $N\to 1$ the gluon-gluon anomalous dimension
behaves like the square root of $\as$.  How can this
behaviour emerge from perturbation theory, which deals only
in integer powers of $\as$?  The answer is that at any
fixed $N\neq 1$ we can expand eq.~(\ref{gamggN}) in a different
way for sufficiently small $\as$:
\beq\label{gamexp}
\gamma_{gg}(N,\as) = \frac{C_A\as}{\pi(N-1)}-
2\left(\frac{C_A\as}{\pi}\right)^2 \frac{1}{(N-1)^3}
+\cdots \;.
\eeq
This series displays the terms that are most singular as
$N\to 1$ in each order.  These terms have been resummed
in the expression (\ref{gamggN}), allowing the perturbation
series to be analytically continued outside its circle of
convergence $|\as|<(\pi/8C_A)|N-1|^2$.  By definition, the
behaviour outside this circle (in particular, at $N=1$)
cannot be represented by the series any more, even though it
is fully implied by it.

At sufficiently small $x$, the $N\to 1$ singularity of the
gluon-gluon anomalous dimension dominates in all
fragmentation functions, and this in turn determines the
asymptotic behaviour of the single-particle inclusive distributions
$F^h$ in eq.~(\ref{Fhxs}). To predict this behaviour quantitatively
we need to take account of the running of $\as$, which can be
done by writing eq.~(\ref{DNtlt}) in the form
\beq
\tilde D(N,s) \sim \exp\left[\int^{s} \gamma_{gg}(N,\as)
\frac{ds'}{s'}\right] 
\eeq
and noting that $\as$ in the integrand should be
$\as(s')$.  We then use eq.~(\ref{s14e7}) to write
\beq
\int^{s} \gamma_{gg}(N,\as(s')) \frac{ds'}{s'}
= -\frac 1 b \int^{\as(s)}\gamma_{gg}(N,\as)\,\frac{d\as}{\as^2}\;,
\eeq
and hence
\beqn\label{DNasy}
\tilde D(N,s)&\propto & \exp\Biggl[
 \frac{1}{b}\sqrt{\frac{2C_A}{\pi\as}} -\frac{1}{4b\as}(N-1)
\nonumber \\
&+& \frac{1}{48b} \sqrt{\frac{2\pi}{C_A\as^3}}(N-1)^2
+\cdots\Biggr]_{\as=\as(s)}\;.
\eeqn

The value of $\tilde D(N,s)$ at  $N=1$ is simply the integral of
the fragmentation function, which gives the {\em average multiplicity},
\beq
 N_i^h(s) \sim \tilde D_i^h(1,s) \propto
\exp\left[\frac{1}{b}\sqrt{\frac{2C_A}{\pi\as(s)}}\right]
\sim \exp\sqrt{\frac{2C_A}{\pi b}\ln s}\;.
\eeq
Thus the average multiplicity of any hadron species should increase
asymptotically faster than any power of $\ln s$ but slower than any
positive power of $s$. Furthermore the relations (\ref{gammas}) imply
that the average multiplicities in gluon and quark jets are
asymptotically in the ratio of their `colour charges' $C_A$
and $C_F$ \cite{Brodsky:1976mg}:
\beq\label{NgNq}
\frac{N_g^h(s)}{N_q^h(s)}\to\frac{C_A}{C_F} = \frac 9 4\;.
\eeq

The behaviour of $\tilde D(N,s)$ near $N=1$
determines the form of small-$x$
fragmentation functions via the inverse Mellin transformation
(\ref{invMell}). Keeping the first three terms in the Taylor
expansion of the exponent, as displayed in
eq.~(\ref{DNasy}), gives a simple Gaussian function of $N$
which transforms into a Gaussian in the variable
$\xi\equiv\ln(1/x)$:
\beq\label{gaussxD}
xD(x,s)\propto \exp\left[-\frac{1}{2\sigma^2}(\xi-
\xi_p)^2\right]\;,
\eeq
where the peak position is
\beq\label{xipeak}
\xi_p = \frac{1}{4b\as(s)}\sim \frac{1}{4}\ln s
\eeq
and the width of the distribution of $\xi$ is
\beq\label{xiwidth}
\sigma = \left(\frac{1}{24b}\sqrt{\frac{2\pi}{C_A\as^3(s)}}
\right)^{\frac{1}{2}} \propto (\ln s)^{\frac{3}{4}}\;.
\eeq

Thus the effect of resummation is to generate a characteristic
hump-backed shape in the variable $\xi =\ln(1/x)$, with a peak
that moves up and expands slowly with increasing $s$.
Including also next-to-leading logarithms, one obtains what
is commonly known as the modified leading-logarithmic approximation
(MLLA) \cite{Azimov:1986by,Fong:1989qy,Dokshitzer:1991wu}.

\section{Hadronization Models}\label{sec:hadro}
\subsection{General ideas}
Before discussing specific models for the hadronization process,
we should review some general ideas that have proved useful in
interpreting hadronization data.
\begin{itemize}
\item {\em Local parton-hadron duality} \cite{Azimov:1985np}.
 Hadronization is long-distance process, involving only small
 momentum transfers. Hence the flows of energy-momentum and
 flavour quantum numbers at hadron level should follow those at
 parton level. Results on inclusive spectra and multiplicities
 support this hypothesis.

\item {\em Universal low-scale $\as$}
\cite{Dokshitzer:1995zt,Dokshitzer:1996ev,Dokshitzer:1996qm}.
 Perturbation theory works well down to low scales, $Q\sim 1$ GeV.
 Assume therefore that
 $\as(Q^2)$ can be defined non-perturbatively for all $Q$, and use it
 in evaluation of Feynman graphs. This approach gives a good description
 of heavy quark spectra and event shapes.
\end{itemize}

\subsection{Specific models}
\begin{figure}\begin{center}
\begin{minipage}{50mm}
\epsfig{file=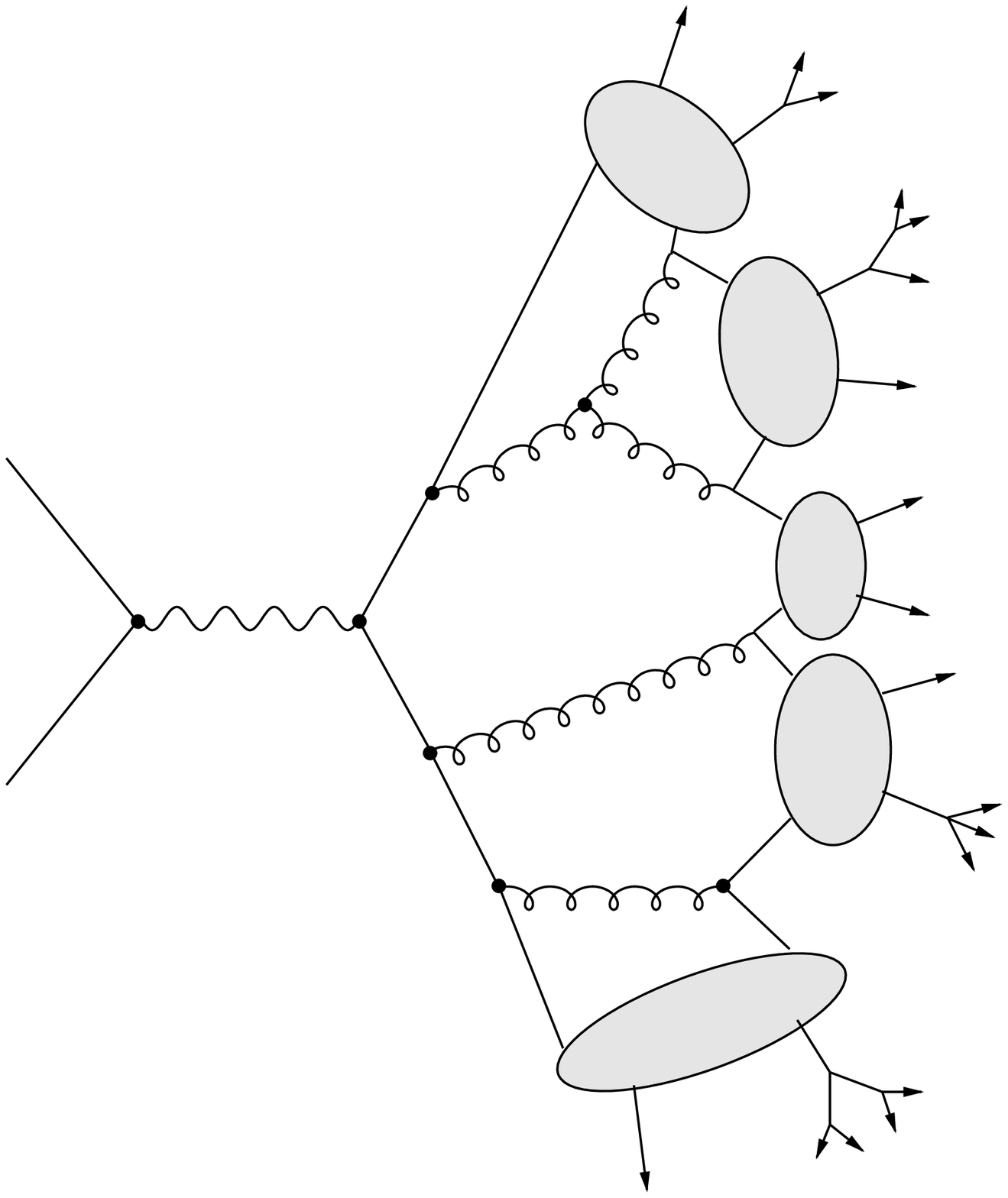,width=5cm}\end{minipage}
\begin{minipage}{50mm}
\epsfig{file=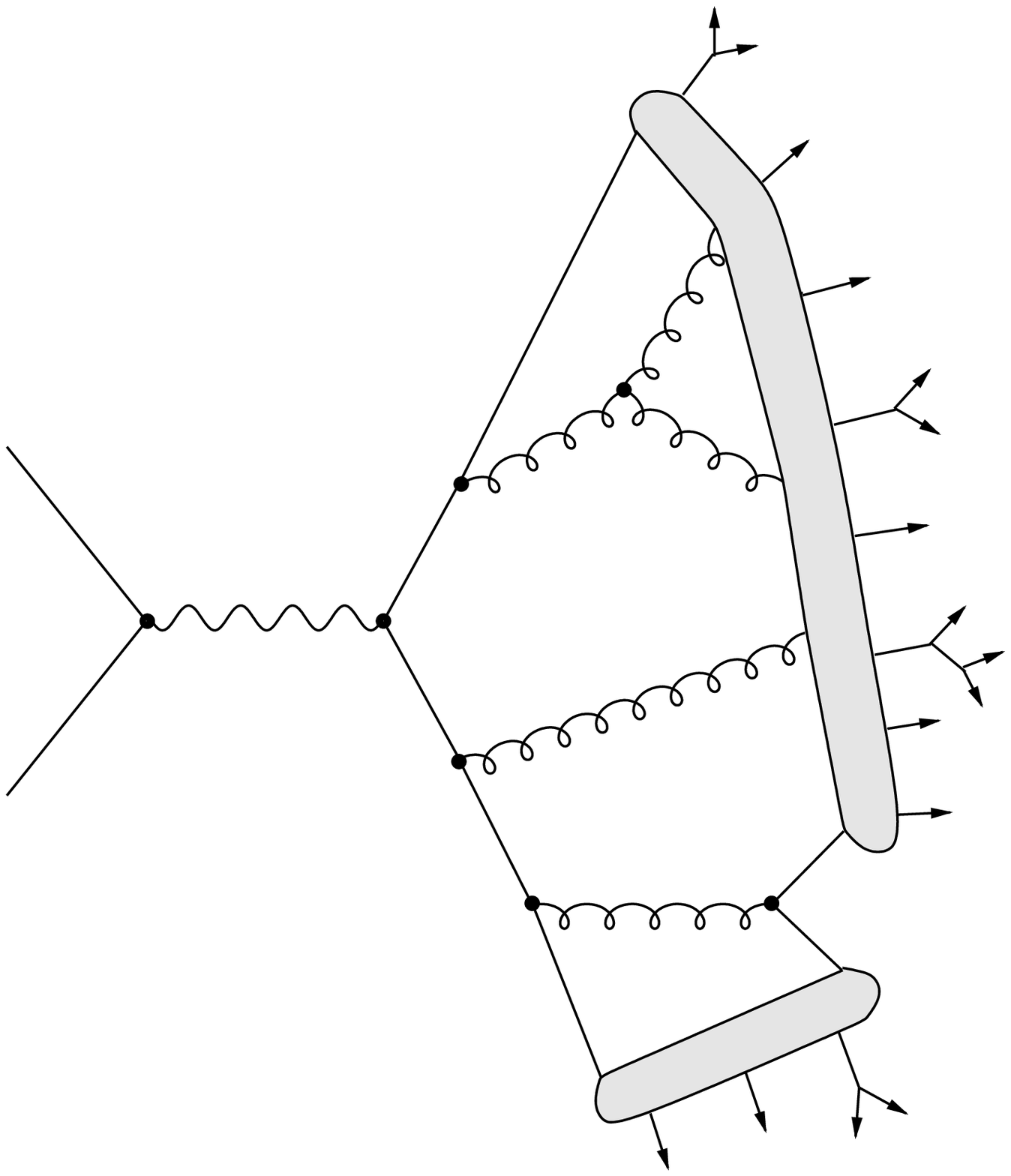,width=5cm}\end{minipage}
\caption{Cluster and string hadronization models.}
\label{fig:clus_string}\end{center}\end{figure}

The above general ideas do not try to describe the mechanism of
hadron formation.  For this we must resort to models.
The main current models are {\em cluster} and {\em string}
hadronization.  We describe briefly the versions used in the
HERWIG and JETSET Monte Carlo event generators, respectively.
In both cases, a {\em parton shower} initiated by the hard process
evolves perturbatively, according to the DGLAP equation, until the scale
of parton virtualities has fallen to some low value $Q_0\sim 1$ GeV,
whereupon the non-perturbative processes assumed
in the model take over (fig.~\ref{fig:clus_string}).
\begin{itemize}
\item {\em Cluster model}
\cite{Marchesini:1984bm}-\cite{Marchesini:1991ch}.
  The model starts with non-perturbative splitting of all
  gluons after the parton shower, $g\to q\bar q$.
  Colour-singlet $q\bar q$ combinations have lower masses and a universal
  spectrum due to the {\em preconfinement}
  \cite{Amati:1979fg,Marchesini:1981cr} property of the shower
  (fig.~\ref{fig:kl_fig9} \cite{Knowles:1997dk}).
  These colour-singlet combinations are assumed to form clusters, which
  mostly undergo simple isotropic decay into pairs of hadrons, chosen
  according to the density of states with appropriate quantum numbers
  \cite{Webber:1984if}.
  This model has few parameters and a natural mechanism for generating
  transverse momenta and suppressing heavy particle production in
  hadronization. However, it has problems in dealing with the decay of
  very massive clusters, and in adequately suppressing baryon and heavy quark
  production.

\begin{figure}\begin{center}
\epsfig{file=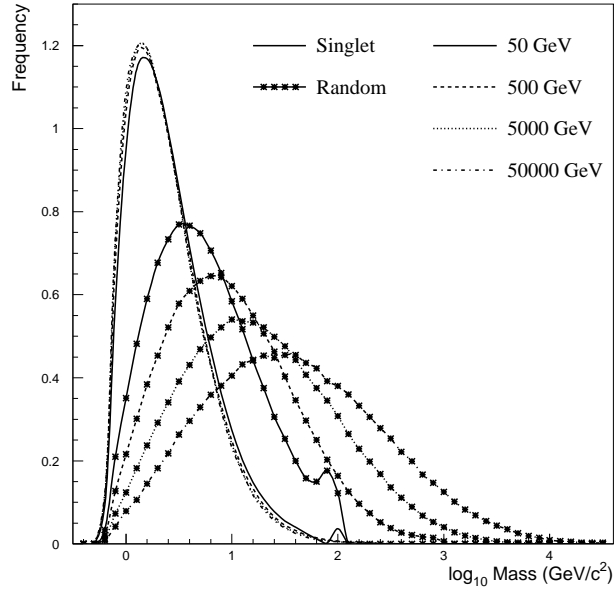,width=8cm}
\caption{Cluster model: mass distribution of $q\bar q$ pairs.}
\label{fig:kl_fig9}\end{center}\end{figure}
\begin{figure}\begin{center}
\epsfig{file=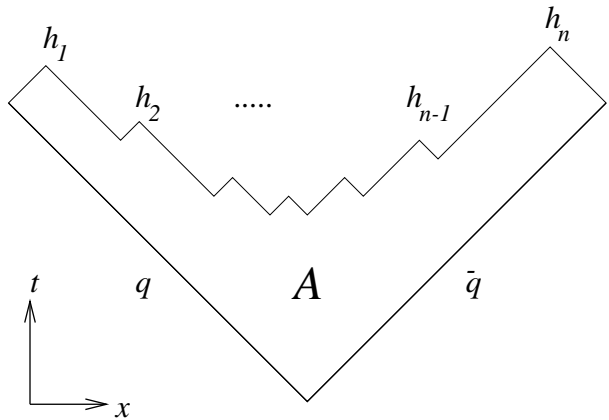,width=8cm}
\caption{String model: space-time picture.}
\label{fig:string_area}\end{center}\end{figure}

\item {\em String model}
\cite{Andersson:1983ia}-\cite{Sjostrand:1994yb}.
  This model is based on the dynamics of a
  relativistic string, representing the colour flux stretched between
  the initial $q\bar q$.  The string produces a linear confinement potential
  and an area law for matrix elements:
  \beq |M(q\bar q\to h_1\cdots h_n)|^2 \propto e^{-bA}\eeq
  where $A$ is the space-time area swept out
  (fig.~\ref{fig:string_area}).  The string breaks up into hadrons via
  $q\bar q$ pair production in its intense colour field.
  Gluons produced in the parton shower give rise to `kinks' on the string.
  The model has extra parameters for the transverse momentum distribution
  and heavy particle suppression. It has some problems describing baryon
  production, but less than the cluster model.

\item {\em The UCLA model} \cite{Buchanan:1987ua,Chun:1993qs}
   is a variant of the JETSET string model which takes the above area law
   for matrix elements more seriously, using
   it to determine the relative rates of production of different hadron
   species. This results in heavy particle suppression without extra
   parameters, the mass-squared
   of a hadron being proportional to its space-time area. At present the
   model still uses extra parameters for $p_T$ spectra, and again has some
   problems describing baryon production.
\end{itemize}

\section{Single-particle yields and spectra}\label{sec:yields}
Tables \ref{tab:proc_mes} and
\ref{tab:proc_bar}\footnote{Updated from ref.~\cite{Knowles:1997dk}.}
compare predictions of the above models\footnote{Recent ALEPH HERWIG
tuning with strangeness suppression 0.8 \cite{Rudolph}.}
with data on Z$^0$ decay from LEP and SLC. Of course, the models have
tunable parameters, but the overall agreement is encouraging. As stated
earlier, the main problems are in the baryon sector, especially for
HERWIG.
\begin{table}\begin{center}{\small
\begin{tabular}{|c|c|c|c|c|c|} \hline 
Particle& Multiplicity& HERWIG& JETSET& UCLA& Expts \\
 & & 5.9 & 7.4 & 7.4 & \\
\hline
Charged       & 20.96(18) & 20.95 & 20.95 & 20.88 &  ADLMO \\
$\pi^\pm$     & 17.06(24) & 17.41 & 16.95 & 17.04 &  ADO \\
$\pi^0$       & 9.43(38)  & 9.97  & 9.59  & 9.61  &  ADLO \\
{\boldmath $\eta$}  & 0.99(4)   & 1.02  & 1.00  & \underline{0.78} & ALO \\
$\rho(770)^0$ & 1.24(10)  & 1.18  & 1.50  & 1.17 & AD \\
$\omega(782)$ & 1.09(9)  & 1.17  & 1.35  & 1.01 & ALO \\
{\boldmath $\eta'(958)$}&0.159(26)& 0.097 & 0.155 & 0.121 & ALO \\
{\boldmath f$_0(980)$}  &0.155(8) & \underline{0.111} &
$\sim$\underline{0.1} & --- & ADO \\
a$_0(980)^\pm$  & 0.14(6) & 0.240 & ---   & --- & O \\
$\phi(1020)$  & 0.097(7)  & 0.104 & \underline{0.194} &\underline{0.132}
& ADO \\
{\boldmath f$_2(1270)$} &0.188(14)& 0.186 &  $\sim 0.2$ & --- & ADO \\
f$_2'(1525)$  & 0.012(6)  & 0.021 & --- & --- & D \\
\hline
K$^\pm$       & 2.26(6)  & 2.16 & 2.30 & 2.24 & ADO \\
{\boldmath K$^0$}   & 2.074(14) & 2.05 & 2.07 & 2.06 & ADLO \\
K$^*(892)^\pm$& 0.718(44) & 0.670 & \underline{1.10} & 0.779 & ADO \\
K$^*(892)^0$  & 0.759(32) & 0.676 & \underline{1.10} & 0.760 & ADO \\
K$_2^*(1430)^0$ & 0.084(40) & 0.111 & --- & --- & DO \\
\hline
D$^\pm$         & 0.187(14) &  \underline{0.276} & 0.174 & 0.196 & ADO \\
D$^0$           & 0.462(26) & 0.506 & 0.490 & 0.497 & ADO \\
D$^*(2010)^\pm$ & 0.181(10) & 0.161 & \underline{0.242} & \underline{0.227}
& ADO \\
D$^\pm_{\rm s}$ & 0.131(20) & 0.115 & 0.129 & 0.130 & O \\
\hline
B$^*$           & 0.28(3)   & 0.201 & 0.260 & 0.254 & D \\
B$^{**}_{\rm u,d}$ & 0.118(24) & \underline{0.013} & ---   & --- & D \\
\hline
J/$\psi$           & 0.0054(4) & \underline{0.0018} & 0.0050 & 0.0050 & ADLO \\
$\psi(3685)$       & 0.0023(5) & 0.0009 & 0.0019 & 0.0019 & DO \\
$\chi_{{\mathrm c}1}$ & 0.0086(27) & \underline{0.0001} & --- & --- & DL \\
\hline
\end{tabular}}
\caption{Meson yields in Z$^0$ decay. Experiments: A=Aleph, D=Delphi,
L=L3, M=Mark II, O=Opal. Bold: new data this year. Underlined: disagreement
with data by more than 3$\sigma$.}
\label{tab:proc_mes}
\end{center}\end{table}
\begin{table}\begin{center}{\small
\begin{tabular}{|c|c|c|c|c|c|} \hline 
Particle& Multiplicity& HERWIG& JETSET& UCLA& Expts \\ 
 & & 5.9 & 7.4 & 7.4 &  \\
\hline
p     & 1.04(4) & \underline{0.863} & \underline{1.19} & 1.09 &  ADO \\
\hline
$\Delta^{++}$ & 0.079(15) & \underline{0.156} & \underline{0.189}
&\underline{0.139} & D \\
              & 0.22(6)   & 0.156 & 0.189 & 0.139 & O \\
\hline
{\boldmath $\Lambda$} & 0.399(8)& 0.387 & 0.385 & 0.382 & ADLO \\
{\boldmath$\Lambda(1520)$}  & 0.0229(25) & --- & --- & --- & DO \\
\hline
$\Sigma^\pm$  & 0.174(16) & 0.154 & 0.140 & 0.118 & DO \\
$\Sigma^0$    & 0.074(9) & 0.068  & 0.073 & 0.074 & ADO \\
$\Sigma^{\star\pm}$ & 0.0474(44) & \underline{0.111} & \underline{0.074}
& \underline{0.074} &
 ADO \\
\hline
$\Xi^-$       & 0.0265(9) & \underline{0.0493} & 0.0271 & \underline{0.0220}
& ADO \\
$\Xi(1530)^0$ & 0.0058(10) & \underline{0.0205} & 0.0053 & 0.0081 & ADO \\
\hline
$\Omega^-$    & 0.0012(2) & \underline{0.0056} & 0.00072 & 0.0011 & ADO \\
\hline
$\Lambda_{\rm c}^+$ & 0.078(17) & \underline{0.0123}  & 0.059 &
\underline{0.026} & O \\
\hline
\end{tabular}}
\caption{Baryon yields in Z$^0$ decay. Legend as in table 1.}
\label{tab:proc_bar}
\end{center}\end{table}

\begin{figure}\begin{center}
\epsfig{file=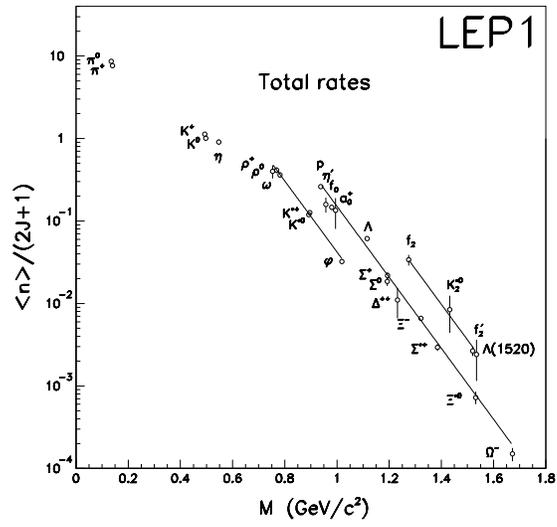,width=8cm}
\caption{Particle yields in Z$^0$ decay.}
\label{fig:chliap}\end{center}\end{figure}

It is remarkable that most measured yields (except for
the $0^-$ mesons, which have special status as Goldstone
bosons) lie on the family of curves
\beq\VEV{n} = a (2J+1) e^{-M/T}\eeq
where $M$ is the mass and $T\simeq 100$ MeV
(fig.~\ref{fig:chliap} \cite{Chliapnikov:1999qi}).
This suggests that mass, rather than quantum numbers, is
the primary factor in determining production rates. Note that,
surprisingly, the orbitally-excited $J=\thrhf$ baryon
$\Lambda(1520)$ (not yet included in models) is produced
almost as much as the unexcited $J=\thrhf$ baryon $\Sigma(1385)$
\cite{Alexander:1996qj,3_147}.

\begin{figure}\begin{center}
\epsfig{file=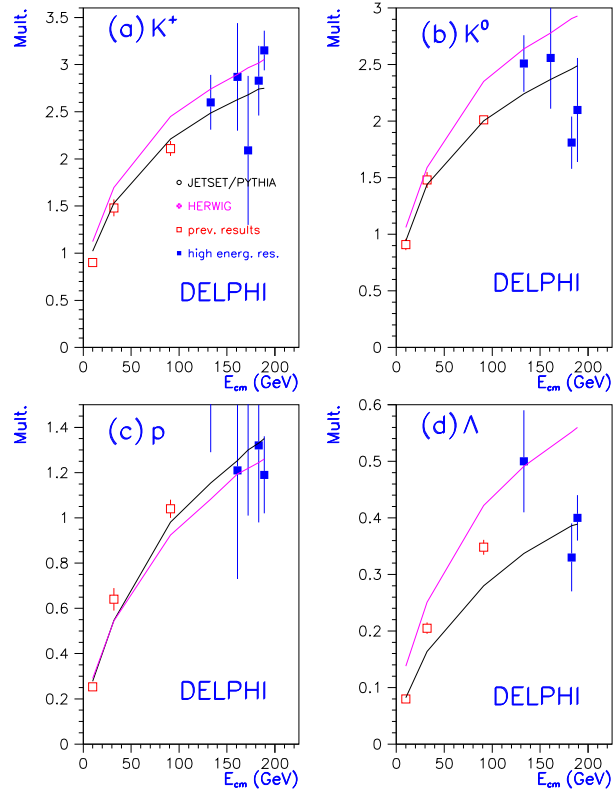,width=8cm}
\caption{Particle yields in $\ee$ annihilation.}
\label{fig:1_229_16}\end{center}\end{figure}

\begin{figure}\begin{center}
\begin{minipage}{60mm}
\epsfig{file=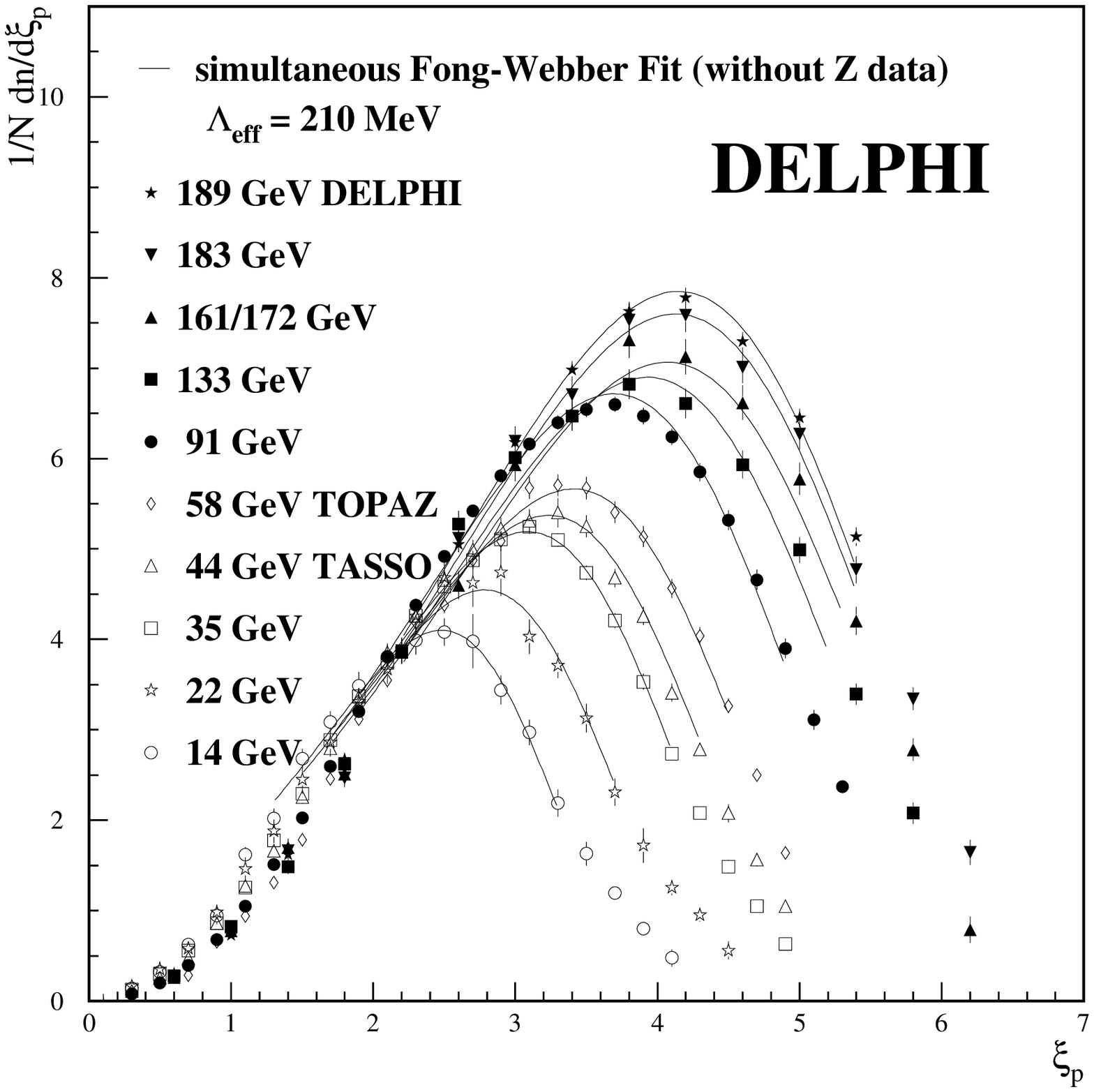,width=60mm}
\end{minipage}
\begin{minipage}{60mm}
\epsfig{file=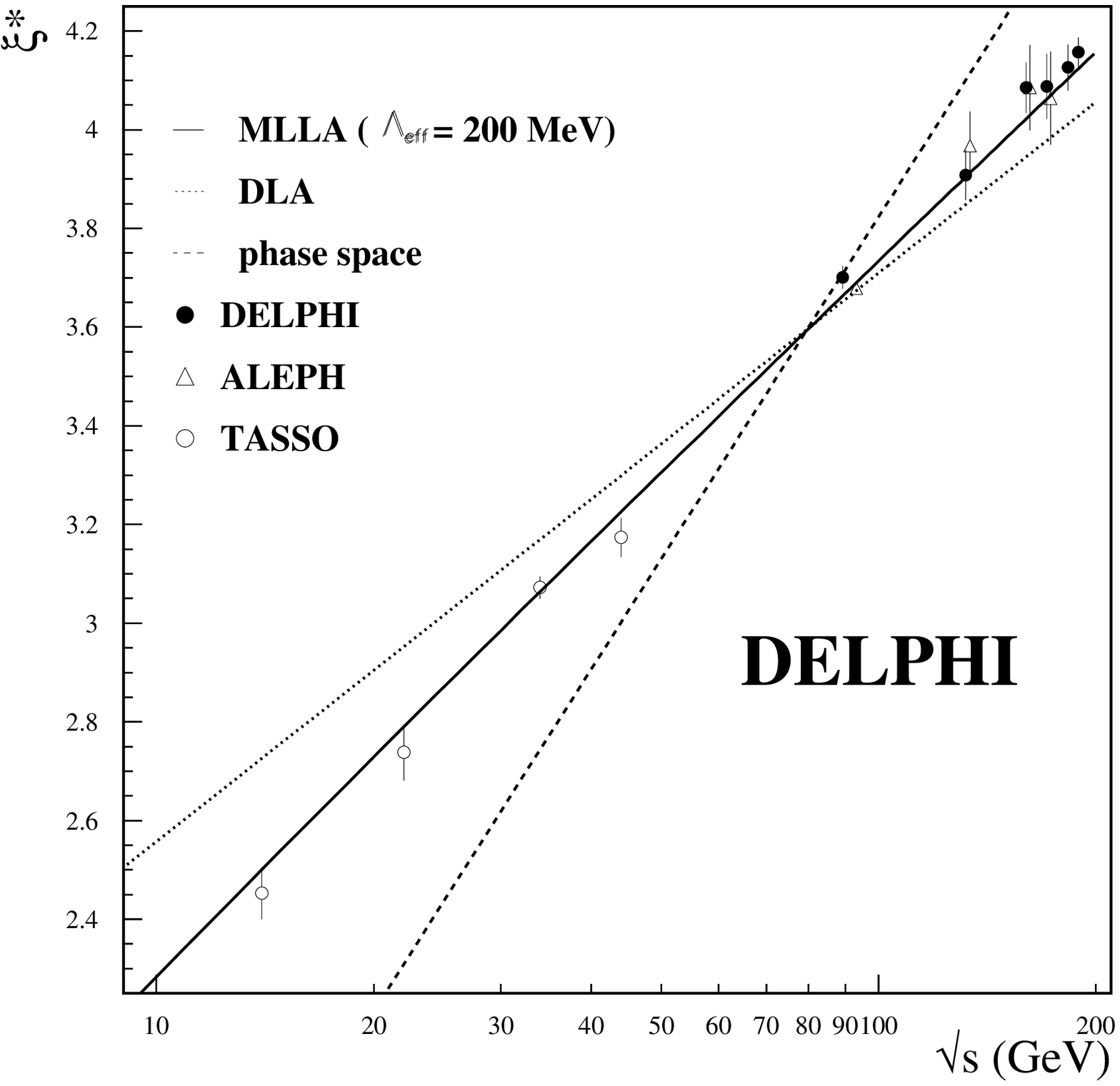,width=60mm}
\end{minipage}
\caption{Low-$x$ fragmentation in $\ee$ annihilation.}
\label{fig:1_225_2}\end{center}\end{figure}

At other energies, model predictions for identified particle yields
are in broad agreement with $\ee$ data (fig.~\ref{fig:1_229_16} \cite{1_229}),
but statistics are of course poorer. Charged particle spectra at low $x$
agree well with the resummed (MLLA) predictions discussed in
sect.~\ref{sec:frag} over a wide  energy range, as illustrated
in fig.~\ref{fig:1_225_2} \cite{1_225}.

\begin{figure}\begin{center}
\epsfig{file=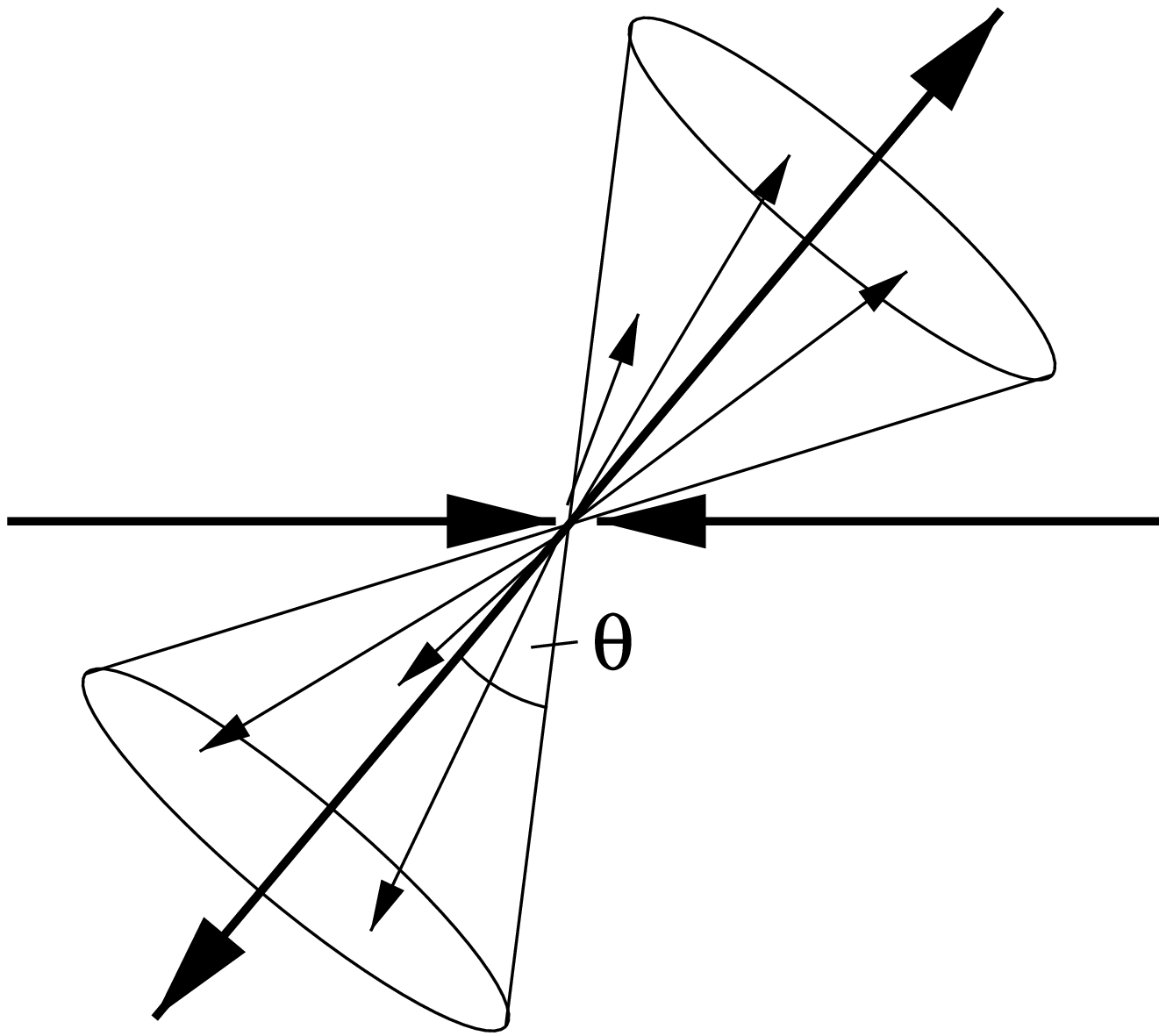,width=60mm}
\caption{Cone angle in  $p\bar p \to$ dijets.}
\label{fig:dijet}\end{center}\end{figure}

\begin{figure}\begin{center}
\begin{minipage}{60mm}
\epsfig{file=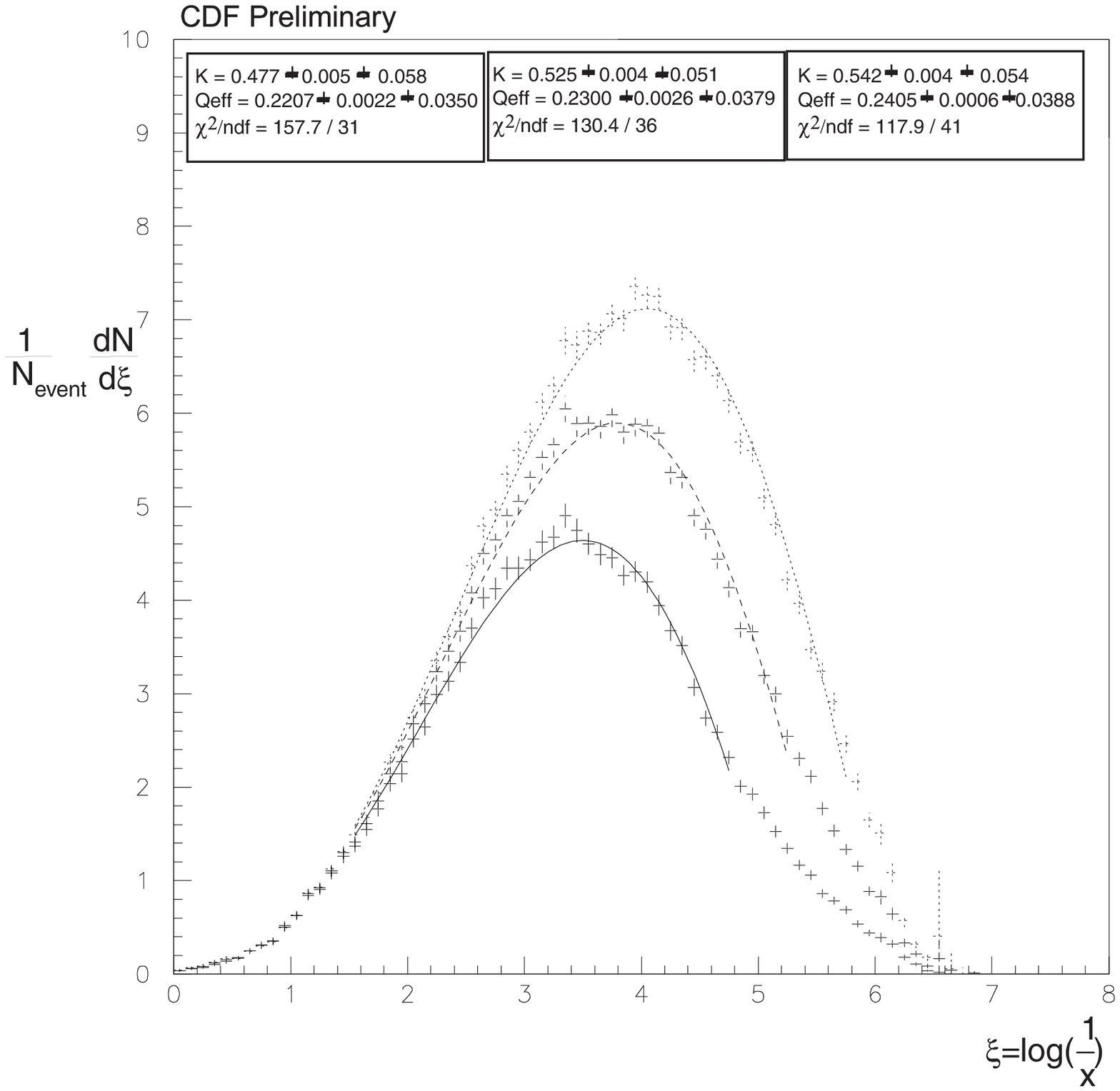,width=55mm}
\end{minipage}
\begin{minipage}{60mm}
\epsfig{file=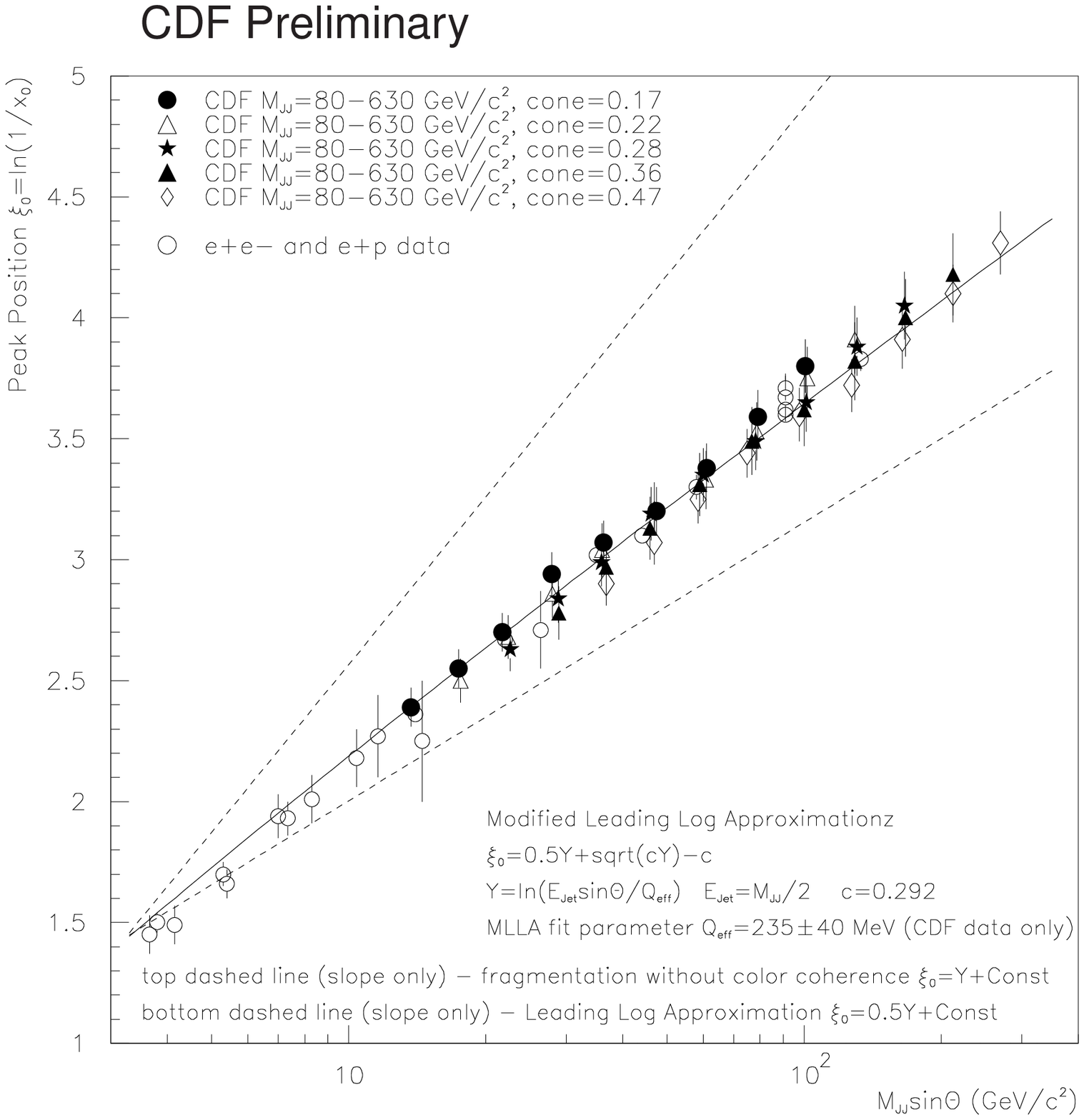,width=55mm}
\end{minipage}
\caption{Low-$x$ fragmentation in $p\bar p \to$ dijets.}
\label{fig:cdf4996_peak_vs_m_new}\end{center}\end{figure}

In $p\bar p \to$ dijets \cite{cdf4996} the relevant scale is taken to be
{$Q=M_{JJ}\sin\theta$} where $M_{JJ}$ is the dijet mass and
$\theta$ is the jet cone angle (fig.~\ref{fig:dijet}). Results
are then in striking agreement with MLLA predictions and with
data from $\ee$ annihilation at $Q=\sqrt s$
(fig.~\ref{fig:cdf4996_peak_vs_m_new}).

\begin{figure}\begin{center}
\epsfig{file=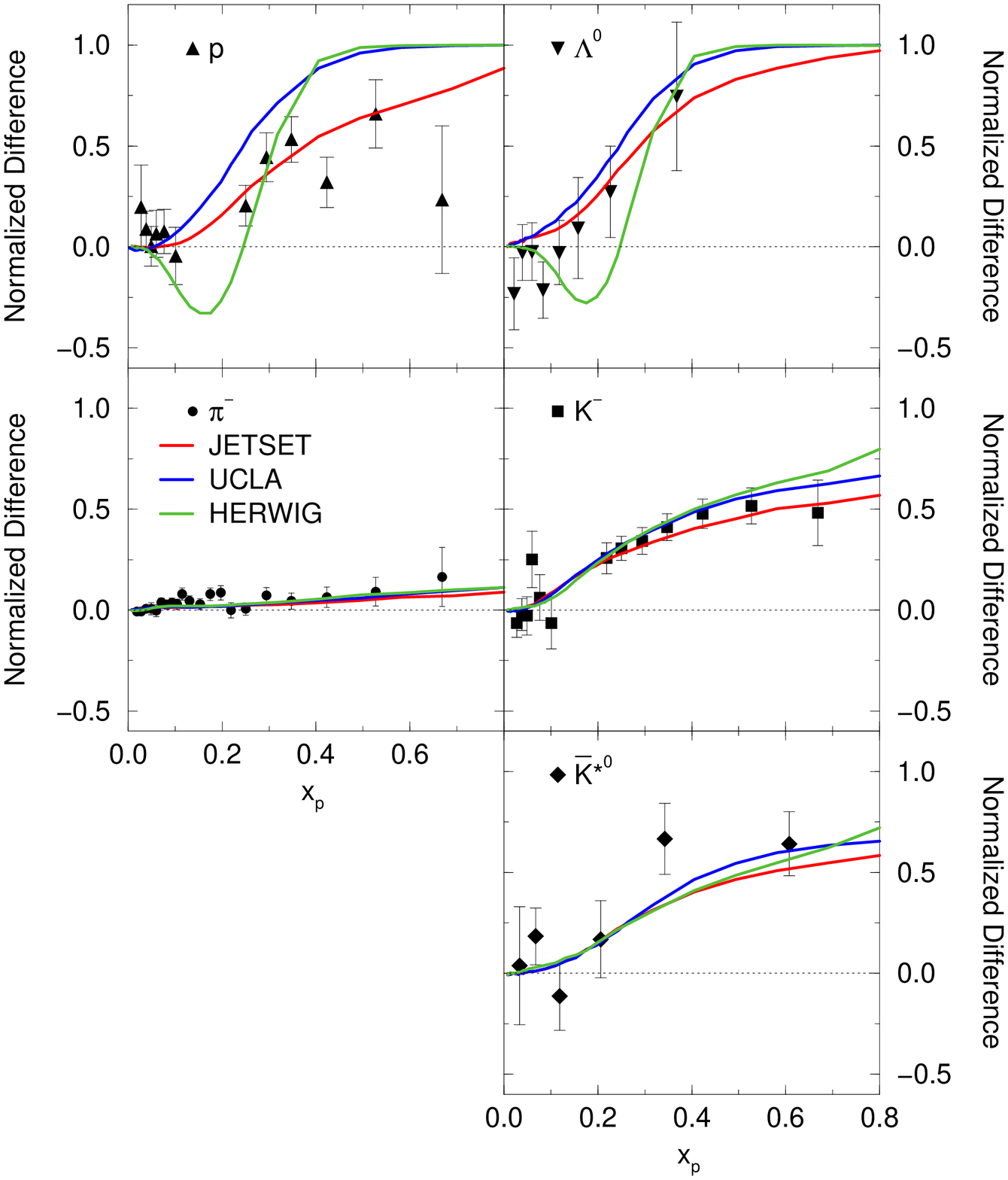,width=70mm}
\caption{Normalized particle--antiparticle differences
in quark jet fragmentation.}
\label{fig:SLD_8159_8}\end{center}\end{figure}

New SLD data include hadron spectra in light quark (rather than antiquark)
fragmentation, selected by hemisphere using the SLC beam polarization
\cite{Abe:1999qh}. One sees strong particle/antiparticle differences
in the expected directions (fig.~\ref{fig:SLD_8159_8}),
bearing in mind the predominance of down-type quarks in Z$^0$ decay.
\begin{figure}\begin{center}
\epsfig{file=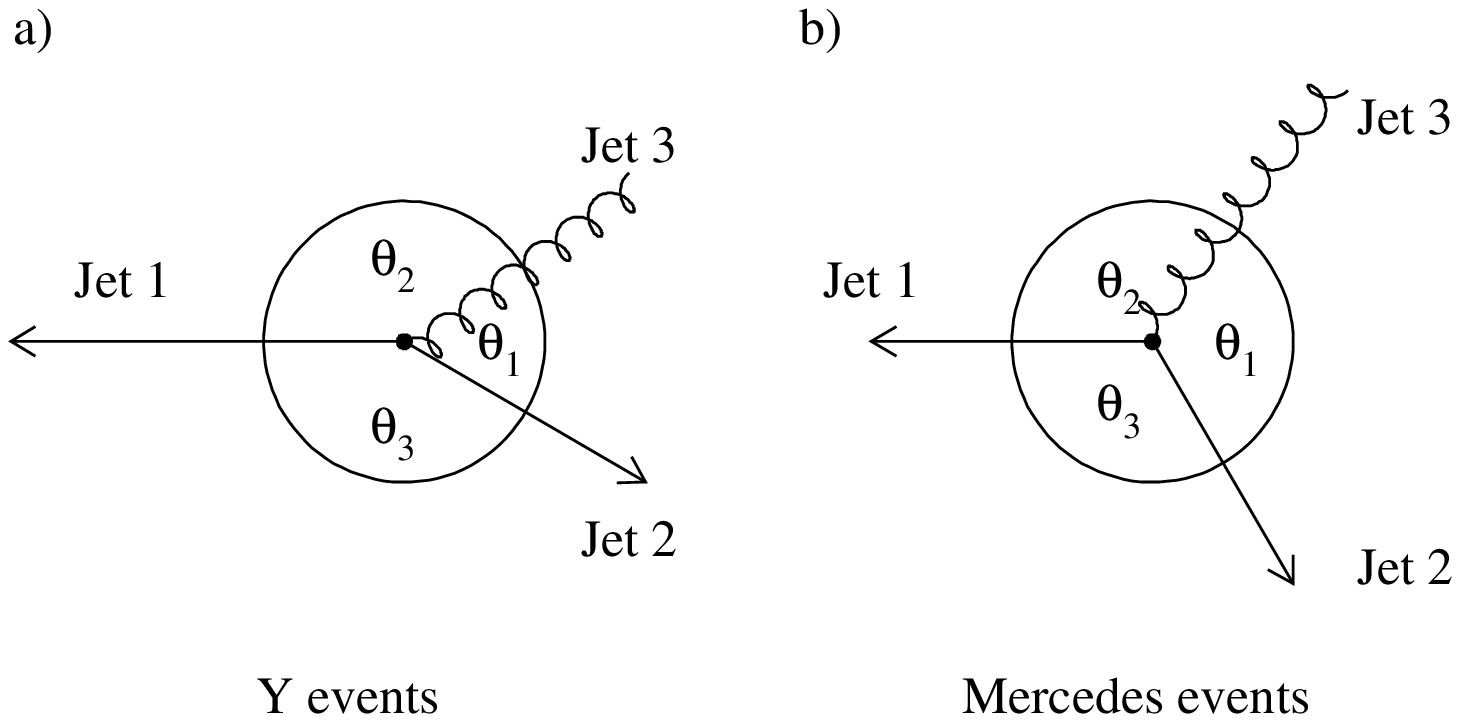,width=80mm}
\caption{Selection of gluon jets by DELPHI.}
\label{fig:Delphi_3_146_1}\end{center}\end{figure}

\section{Quark and gluon jets}\label{sec:qg}
\begin{figure}\begin{center}\epsfig{file=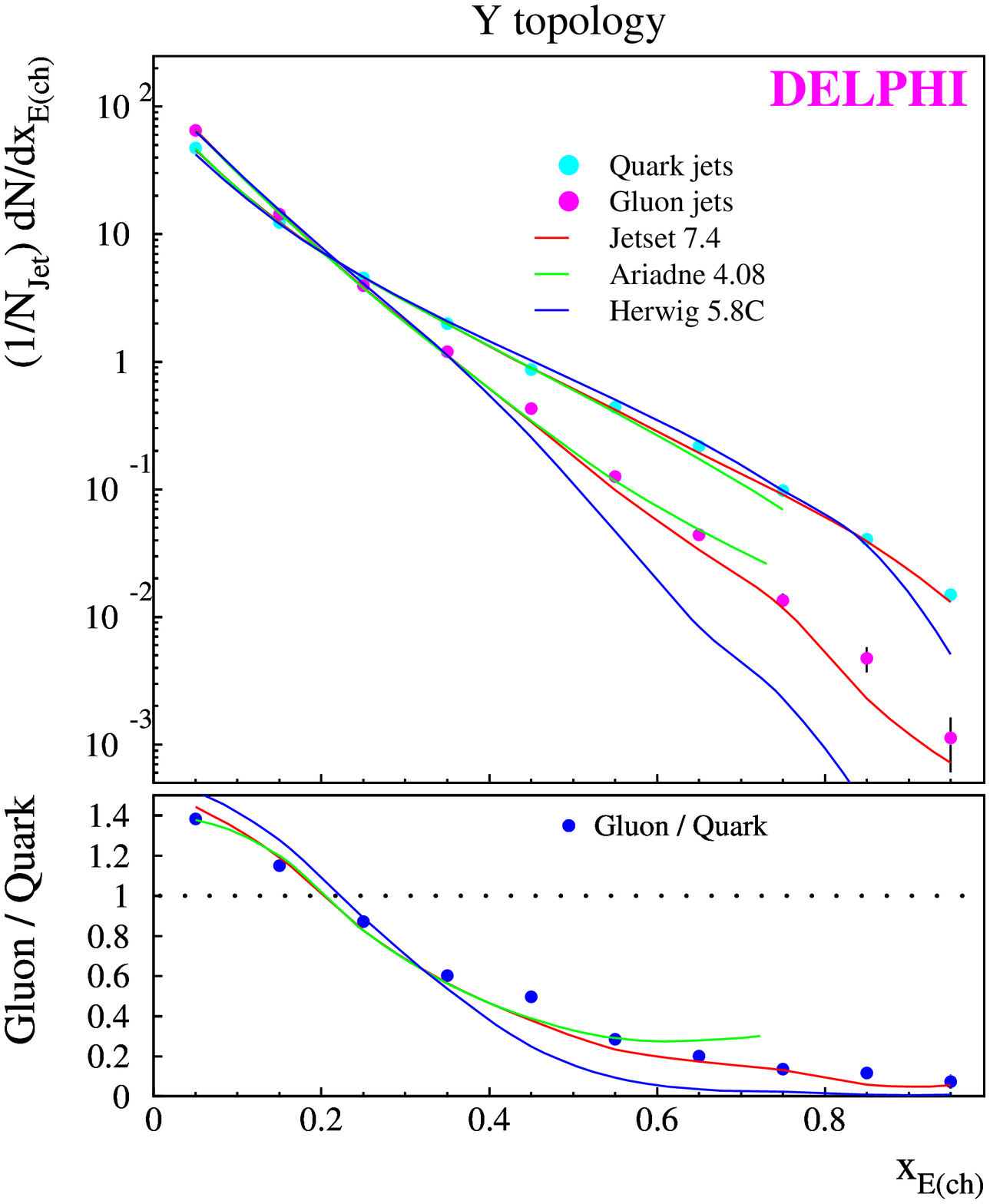,width=70mm}
\caption{Charged particle spectra in quark and gluon jets.}
\label{fig:Delphi_1_571_9}\end{center}\end{figure}

\begin{figure}\begin{center}\epsfig{file=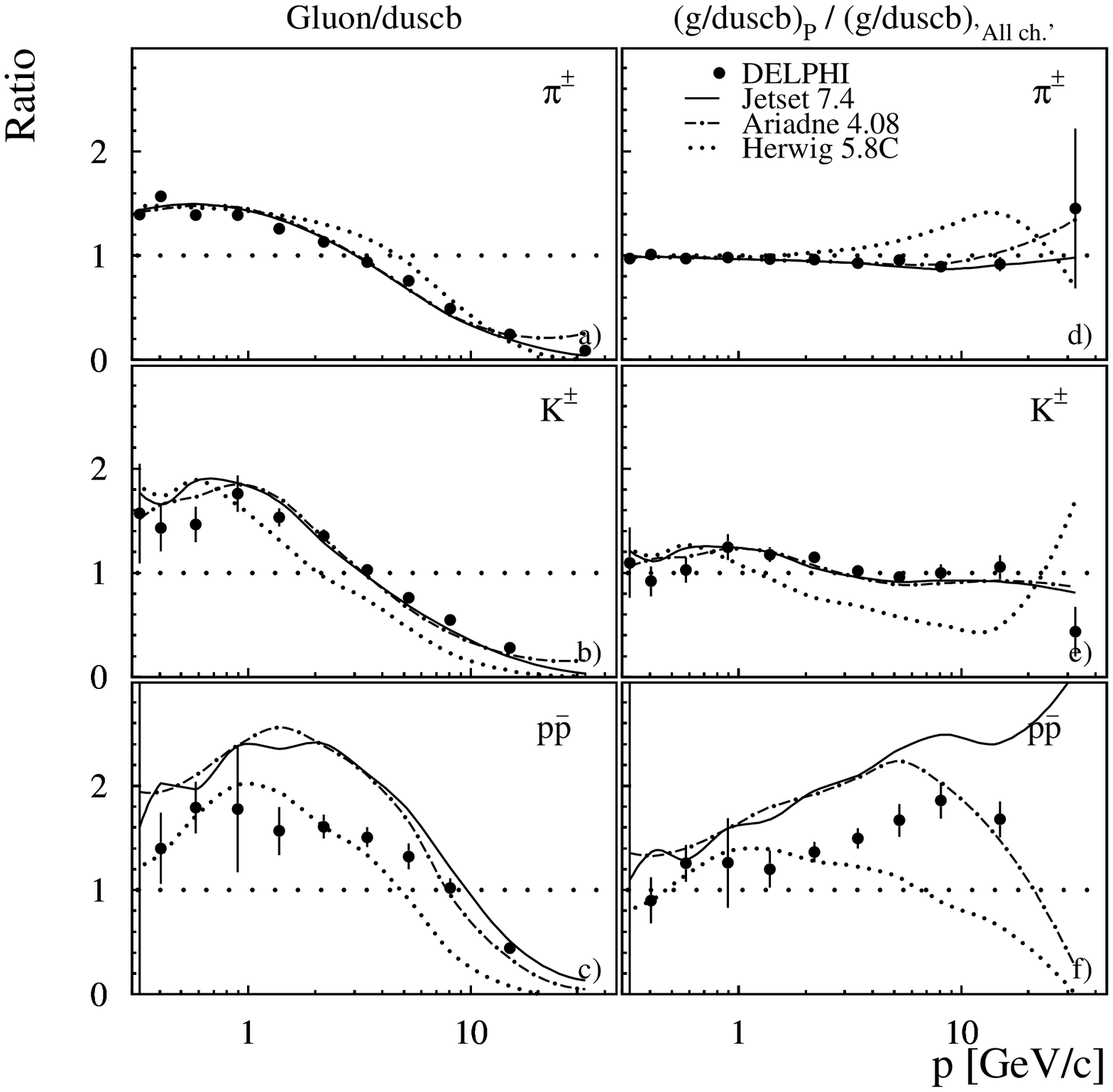,width=80mm}
\caption{Comparisons of particle spectra in quark and gluon jets.}
\label{fig:Delphi_3_146_53}\end{center}\end{figure}

DELPHI \cite{3_146} select gluon jets by anti-tagging
heavy quark jets in `Y' and `Mercedes' three-jet events
(fig.~\ref{fig:Delphi_3_146_1}).  As expected, the higher
colour charge of the gluon ($C_A=3$ vs.\ $C_F=4/3$) leads to
a softer spectrum and higher overall multiplicity
(fig.~\ref{fig:Delphi_1_571_9}).
In general the relative multiplicities of identified particles are
consistent with those of all charged, with no clear excess of
any species in gluon jets (fig.~\ref{fig:Delphi_3_146_53}).
In particular there is no enhanced $\phi(1020)$ or $\eta$ production:

\noindent
DELPHI \cite{3_146}: $N_g(\phi)/N_q(\phi) = 0.7\pm 0.3$\\
OPAL \cite{1_4}: $N_g(\eta)/N_q(\eta) = 1.29\pm 0.11$


\begin{figure}\begin{center}
\begin{minipage}{60mm}
\epsfig{file=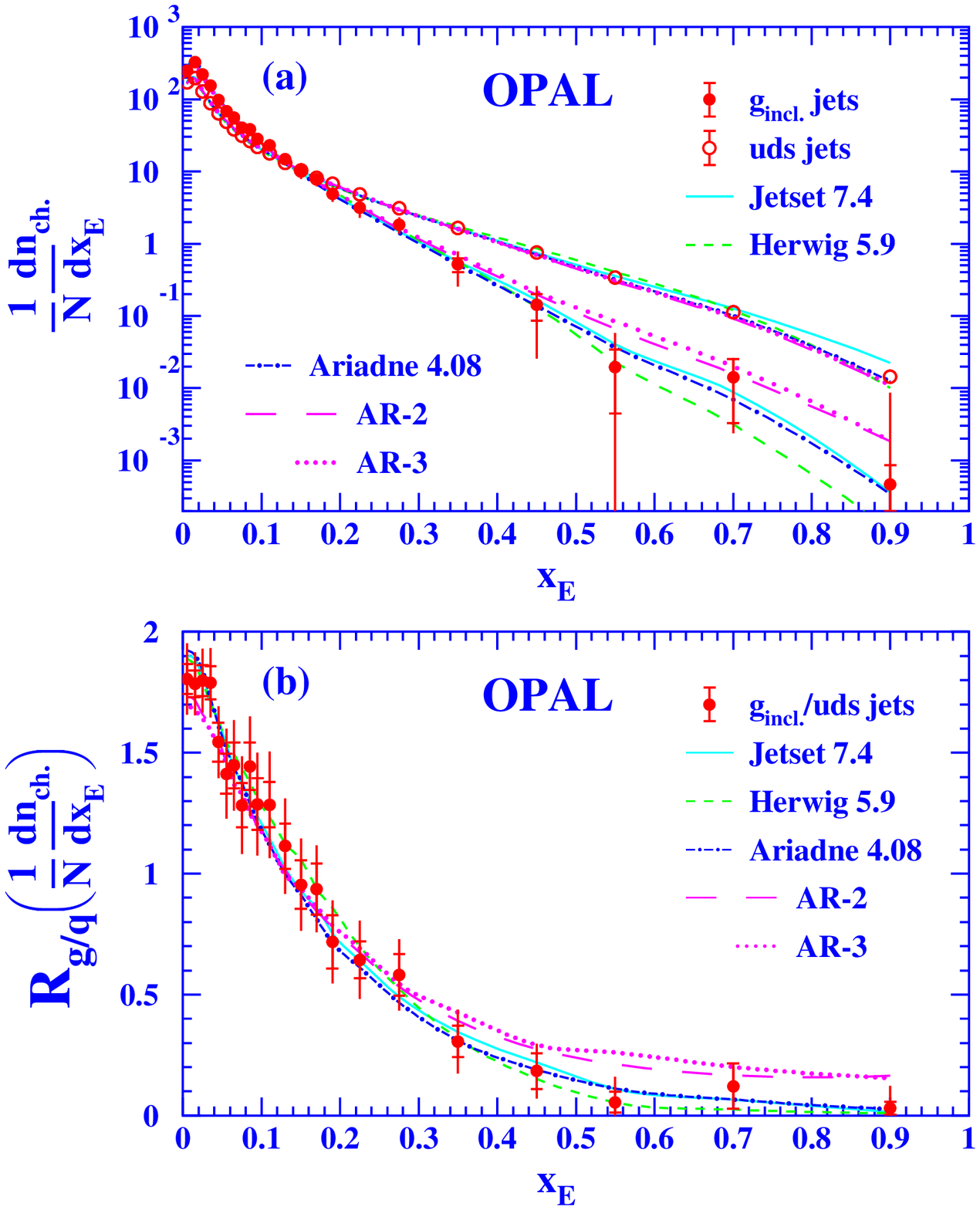,width=60mm}
\end{minipage}
\begin{minipage}{60mm}
\epsfig{file=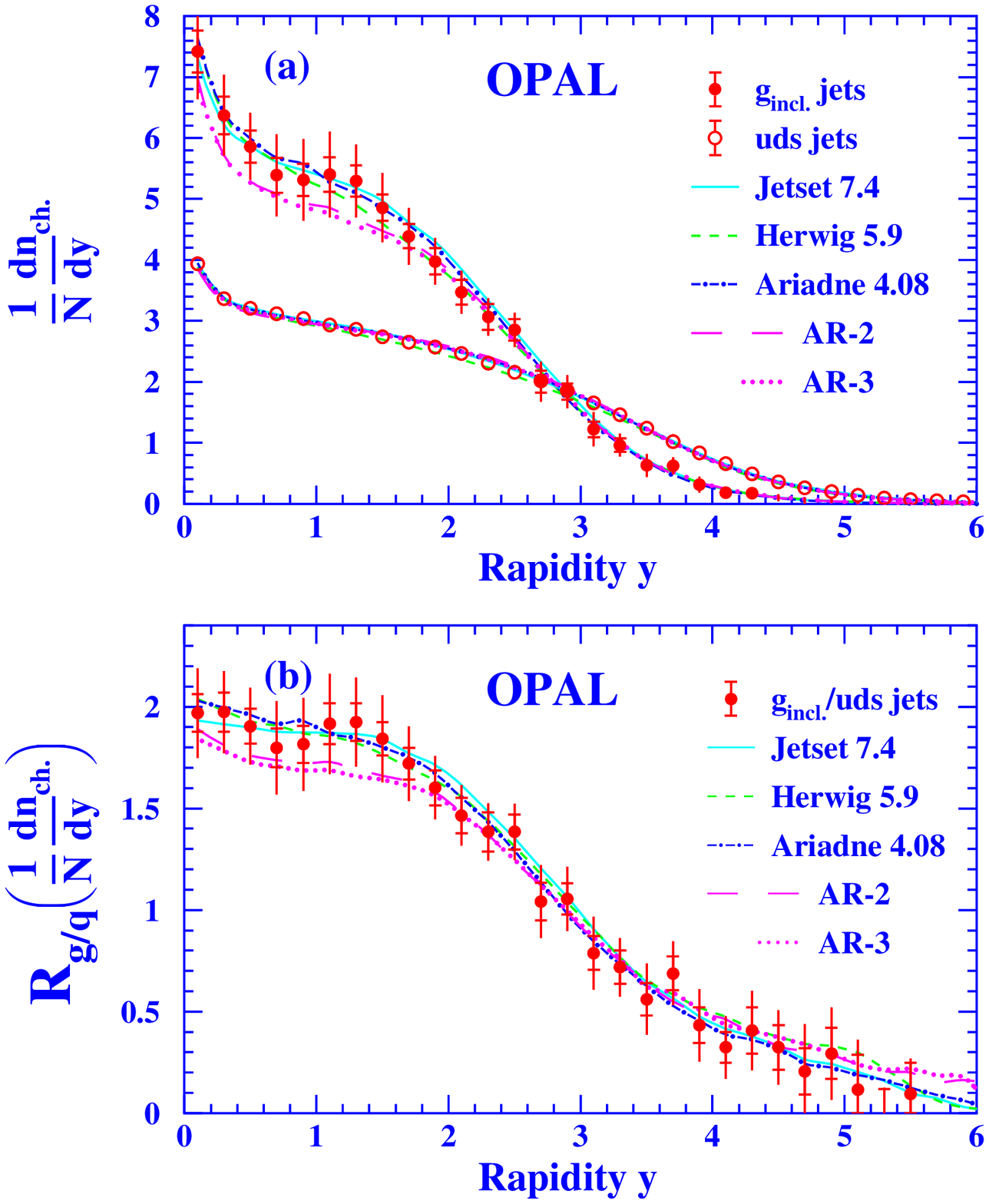,width=60mm}
\end{minipage}
\caption{Momentum fraction and rapidity distributions in quark
and gluon jets.}\label{fig:Opal_24_5}\end{center}\end{figure}

OPAL \cite{Abbiendi:1999pi} select gluon jets recoiling against two
tagged $b$-jets in the same hemisphere. Monte Carlo studies indicate
that such jets should be similar to those emitted by a point source of
gluon pairs. The qualitative message from the data is again clear
(fig.~\ref{fig:Opal_24_5}): Gluon jets have softer fragmentation
than light quark jets, and higher multiplicity.
The precision of the data is now such that
next-to-leading order calculations of the relevant
coefficient functions, taking into account the experimental
selection procedures, are needed to check universality
of the extracted gluon fragmentation function.

The ratio of gluon/quark multiplicities at low rapidity (large
angle) is close to the ratio of colour charges
{$r\equiv C_A/C_F=2.25$},
in agreement with local parton-hadron duality:
$$\mbox{OPAL: }
{r_{ch}(|y|<1) = 1.919\pm 0.047\pm 0.095}\;.$$
According to eq.~(\ref{NgNq}), the overall multiplicity
ratio should also approach the value $C_A/C_F$ asymptotically, but
at present energies the contribution from higher rapidities
is substantial and this leads to a smaller ratio.

Monte Carlo studies \cite{Abbiendi:1999pi}
suggest that a better measure of $C_A/C_F$ 
is obtained by selecting low-momentum hadrons with relatively
large transverse momentum (i.e.\ low rapidity).  This gives
$$\mbox{OPAL: }
r_{ch}(p<4,\,0.8<p_T<3\;\mbox{GeV}) = 2.29\pm 0.09\pm 0.015\;.$$

\begin{figure}\begin{center}
\begin{minipage}{60mm}
\epsfig{file=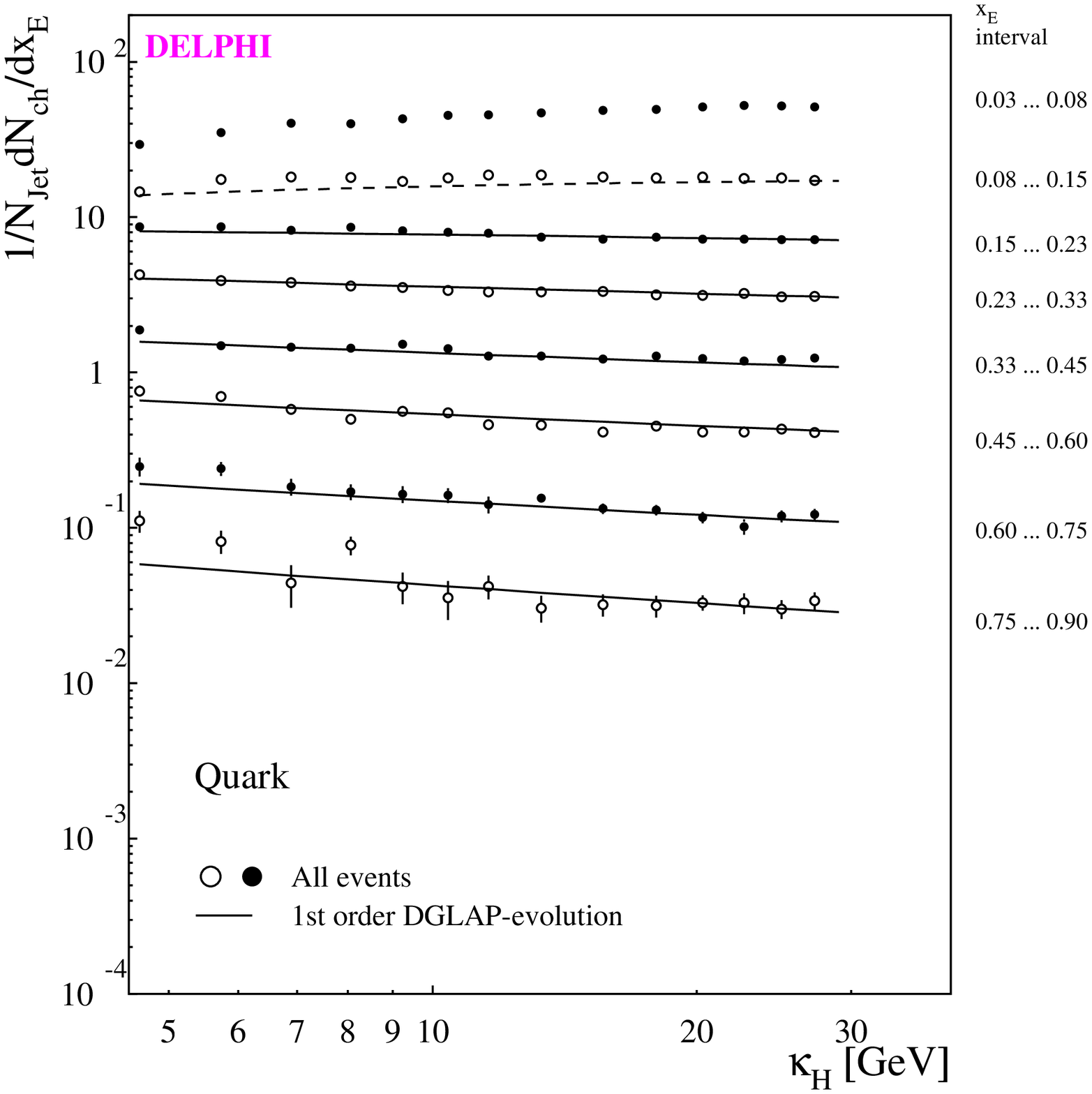,width=55mm}\end{minipage}
\begin{minipage}{60mm}
\epsfig{file=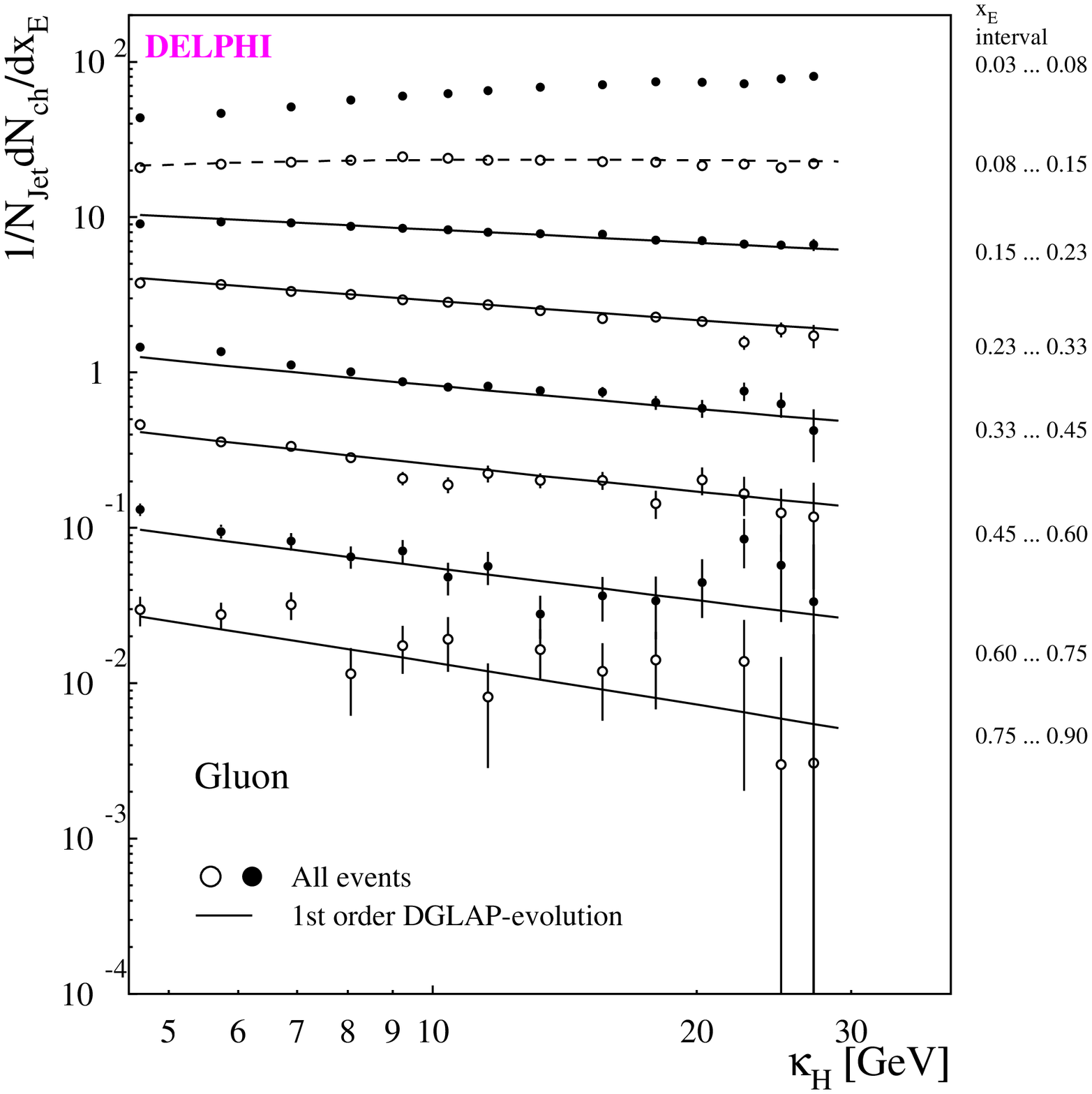,width=55mm}\end{minipage}
\caption{Scale dependence of quark and gluon fragmentation.}
\label{fig:Delphi_1_571_16}\end{center}\end{figure}

\begin{figure}\begin{center}\epsfig{file=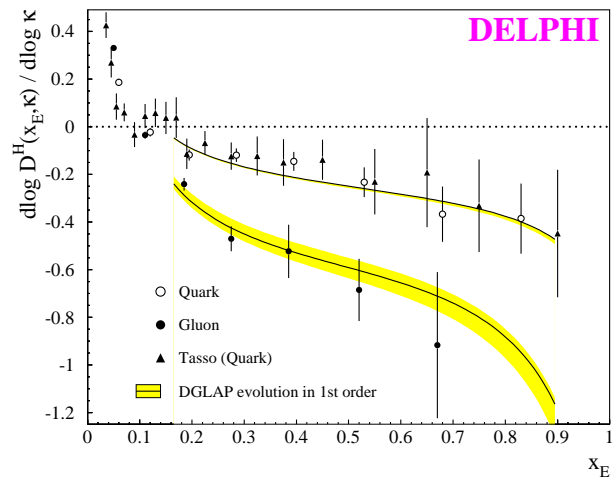,width=85mm}
\caption{Logarithmic gradients of quark and gluon fragmentation.}
\label{fig:Delphi_1_571_17}\end{center}\end{figure}

DELPHI \cite{1_571} have observed scaling violation in quark and gluon jet
fragmentation separately (fig.~\ref{fig:Delphi_1_571_16})
by studying the dependence on the scale
\beq \kappa_H = \Ejet\,\sin (\theta/2)\;\simeq\;\half\sqrt{sy_3}\eeq
where $\theta$ is the angle to the closest jet and
$y_3$ is the Durham jet resolution \cite{Catani:1991hj}
at which 3 jets are just resolved.  This is expected
to be the relevant scale when $y_3$ becomes small.
One sees clearly that there is more scaling violation in gluon jets
(fig.~\ref{fig:Delphi_1_571_17}). The ratio provides another
measure of $C_A/C_F$:
$$\mbox{DELPHI: }
r_{\mbox{\scriptsize sc.viol.}} = 2.23\pm 0.09\pm 0.06\;.$$


A crucial point in the DELPHI analysis is that 3-jet events are not
selected using a fixed jet resolution {$\ycut$}, but rather
each event is clustered to precisely 3 jets.  This avoids `biasing'
the gluon jet sample by preventing further jet emission above $\ycut$.

\begin{figure}\begin{center}\epsfig{file=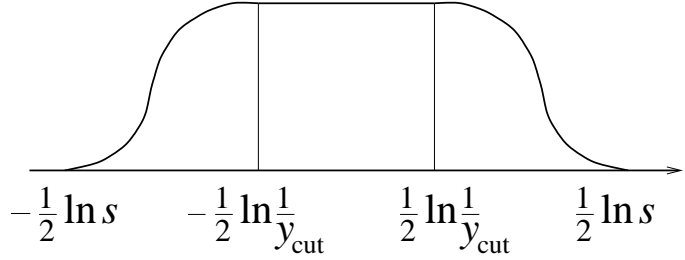,width=90mm}
\caption{Rapidity plateau in two-jet events.}
\label{fig:twojet}\end{center}\end{figure}

\begin{figure}\begin{center}\epsfig{file=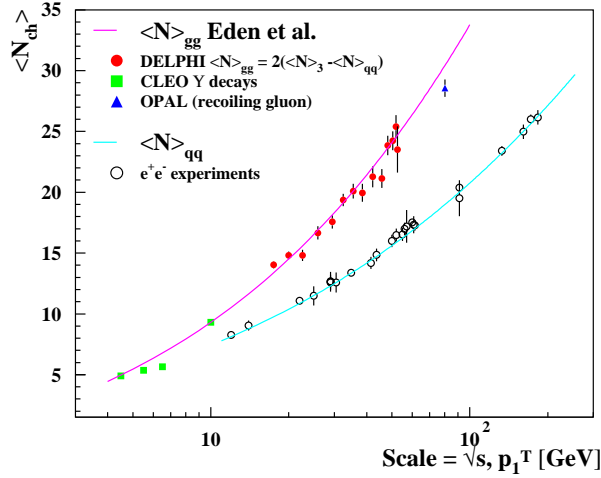,width=80mm}
\caption{Average $q\bar q$ and $gg$ multiplicities deduced from 2- and 3-jet
events.}\label{fig:Delphi_extra_ngg}\end{center}\end{figure}

The same point is well illustrated in analyses of average
multiplicities in 2- and 3-jet events
\cite{Catani:1992tm,Eden:1999vc,Abreu:1999rs}.
If $N_{q\bar q}(s)$ is the `unbiased' $q\bar q$ multiplicity,
then in events with precisely 2 jets at resolution $\ycut$ there is
a rapidity plateau of length $\ln(1/\ycut)$ (see fig.~\ref{fig:twojet})
and the multiplicity is
\beq
N_2(s,\ycut) \simeq N_{q\bar q}(s\ycut)
+\ln(1/\ycut)N'_{q\bar q}(s\ycut)
\eeq
where $N'(s)\equiv sdN/ds$.  Clustering each event to 3 jets we get this
multiplicity with {$y_3$} in place of $\ycut$, plus an unbiased gluon jet:
\beq
N_3(s) \simeq N_2(s,y_3) +\half N_{gg}(sy_3)\;.
\eeq
Thus one can extract the unbiased $gg$ multiplicity,
plotted in fig.~\ref{fig:Delphi_extra_ngg}
vs. $p_1^T\sim \sqrt{sy_3}$ \cite{Delphi_extra}.
The ratio of $gg/q\bar q$ slopes gives yet another measure of $C_A/C_F$
\cite{Abreu:1999rs}:
$$r_{\mbox{\scriptsize mult}} = 2.246\pm 0.062(\mbox{stat.})
\pm 0.080(\mbox{sys.})\pm 0.095(\mbox{theo.})\;.$$


\section{Current and target fragmentation in DIS}\label{sec:DIS}
\begin{figure}\begin{center}\epsfig{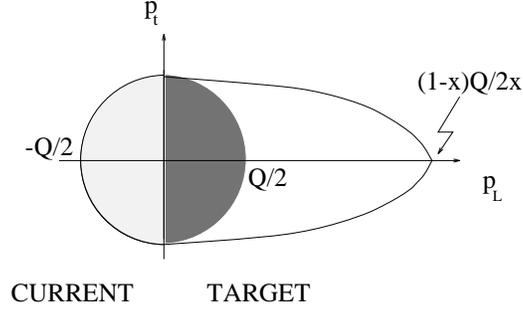}
\caption{Breit frame current and target regions in DIS.}
\label{fig:Zeus_537_1}\end{center}\end{figure}

The H1 \cite{Adloff:1997fr} and ZEUS \cite{Breitweg:1999nt} experiments
at HERA have studied the distributions of $x_p=2|\bom p|/Q$ in the current
and target hemispheres in the {\em Breit frame of reference}.
The Breit frame is the one in which
the 4-momentum of the virtual photon exchanged in DIS lies entirely
along the negative $z$-axis, $q^\mu = (0,0,0,-Q)$, while the target
proton has $P^\mu = (Q,0,0,Q)/2x$ (neglecting the proton mass), where
$x=-q^2/2p\cdot q$ is the Bjorken variable. In this frame, to zeroth
order in $\as$, the virtual photon simply strikes a constituent of
the target with momentum $Q/2$ and reverses its momentum.  The
remnant of the target is then left with momentum $(1-x)Q/2x$
(fig.~\ref{fig:Zeus_537_1}).

\begin{figure}\begin{center}\epsfig{file=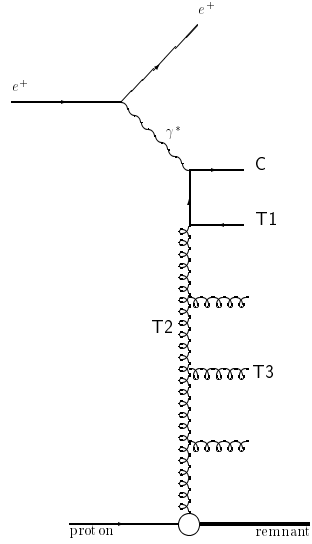,width=40mm}
\caption{Contributions to the final state in DIS.}
\label{fig:Zeus_537_2}\end{center}\end{figure}

In higher orders one expects the current hemisphere to contain
fragmentation products of the current
jet (C in fig.~\ref{fig:Zeus_537_2}), similar to half an $\ee$ event.
In the target hemisphere, the contribution T1 is
similar to C, T2 gives extra particles with $x_p<1$, while T3 gives
$x_p\gtap 1$, generally outside detector acceptance.
\begin{itemize}
\item In the {\em current hemisphere} the charged multiplicity is indeed
similar to $\ee$ (fig.~\ref{fig:Zeus_537_10fit} \cite{Breitweg:1999nt}).
Differences at low $Q^2$ are consistent with the expected
boson-gluon fusion contribution.
The distribution of $\xi=\ln(1/x_p)$ is also
similar to $\ee$, i.e.\ close to Gaussian with little Bjorken $x$
dependence (fig.~\ref{fig:Zeus_537_12}).

\begin{figure}\begin{center}\epsfig{file=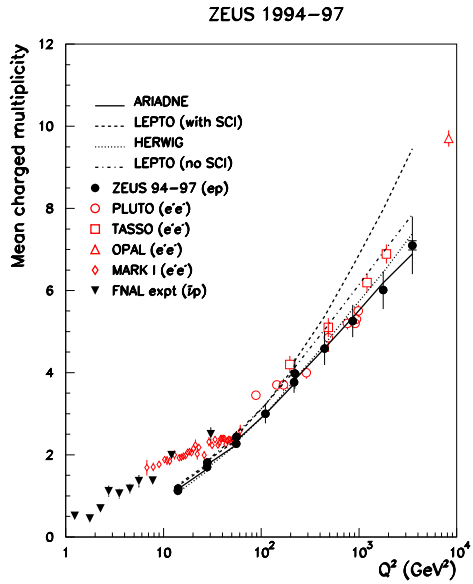,width=8cm}
\caption{Charged multiplicity in current hemisphere.}
\label{fig:Zeus_537_10fit}\end{center}\end{figure}
\begin{figure}\begin{center}\epsfig{file=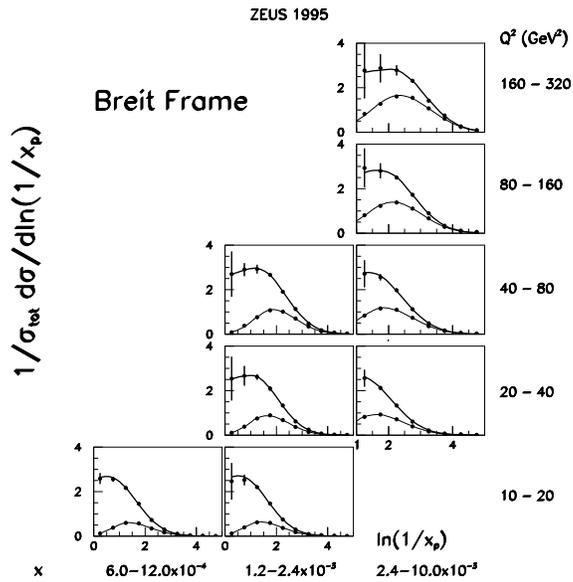,width=80mm}
\caption{Fragmentation in DIS. Upper data (heavy curve) target region,
lower data (light curve) current region.}\label{fig:Zeus_537_12}
\end{center}\end{figure}

At low $Q^2$ there is evidence of strong subleading corrections.
The distribution is skewed towards higher values of
$\xi$ (smaller $x_p$), contrary to MLLA predictions
(fig.~\ref{fig:Zeus_537_4sk}). The quantity plotted is
\beq\mbox{Skewness} \equiv
\VEV{(\xi-\bar\xi)^3}/\VEV{(\xi-\bar\xi)^2}^{\thrhf}\;.
\eeq

\begin{figure}\begin{center}\epsfig{file=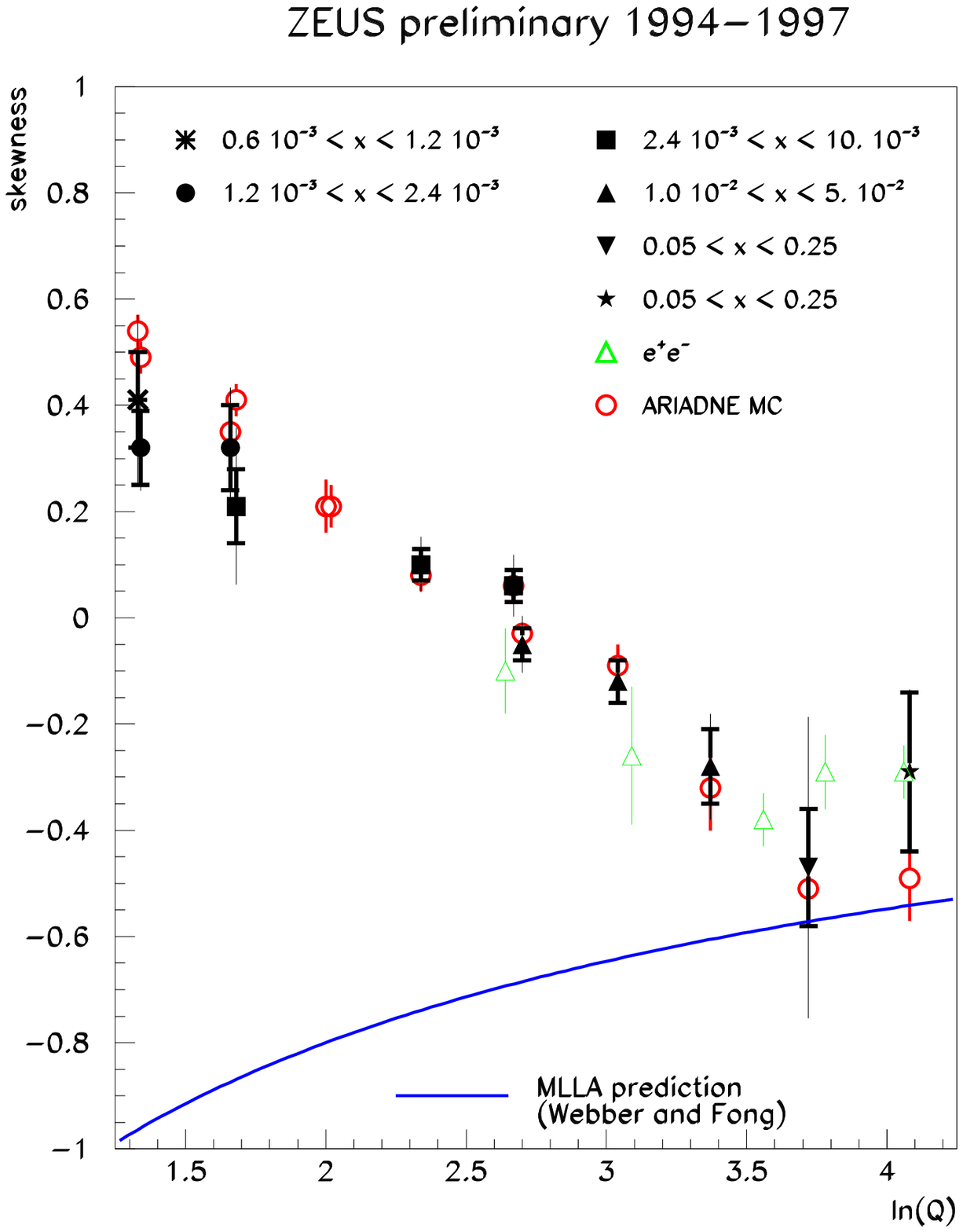,width=60mm}
\caption{Skewness in current fragmentation region.}
\label{fig:Zeus_537_4sk}\end{center}\end{figure}
\begin{figure}\begin{center}\epsfig{file=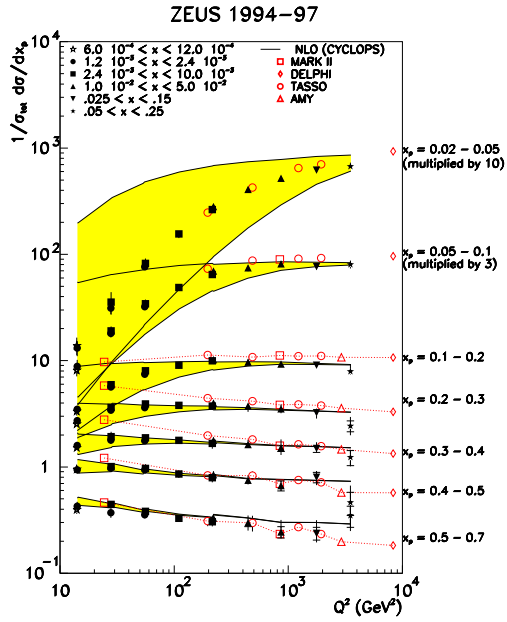,width=70mm}
\caption{Scaling violation in DIS fragmentation.}
\label{fig:Zeus_537_7ee}\end{center}\end{figure}

On the other hand, the data lie well {\em below} the fixed-order
perturbative prediction \cite{Graudenz:1997an} at low $x_p$ and $Q^2$
(fig.~\ref{fig:Zeus_537_7ee}).
Discrepancies could be due to power-suppressed ($1/Q^2$)
corrections, of dynamical and/or kinematical origin.
The bands in fig.~\ref{fig:Zeus_537_7ee}
correspond to an ad-hoc correction factor
\beq
\left[1+\left(\frac{m_{\mbox{\scriptsize eff}}}{Qx_p}\right)^2\right]^{-1}
\qquad (0.1 <m_{\mbox{\scriptsize eff}}< 1 \mbox{ GeV}).\eeq

\item In the {\em target hemisphere} there is also disagreement with
MLLA \cite{Breitweg:1999nt},
possibly due to the T3 contribution ``leaking'' into the region $x_p<1$.
If anything, Monte Carlo models predict too much leakage
(fig.~\ref{fig:Zeus_537_13}).
Little $Q^2$ dependence is evident.
\begin{figure}\begin{center}\epsfig{file=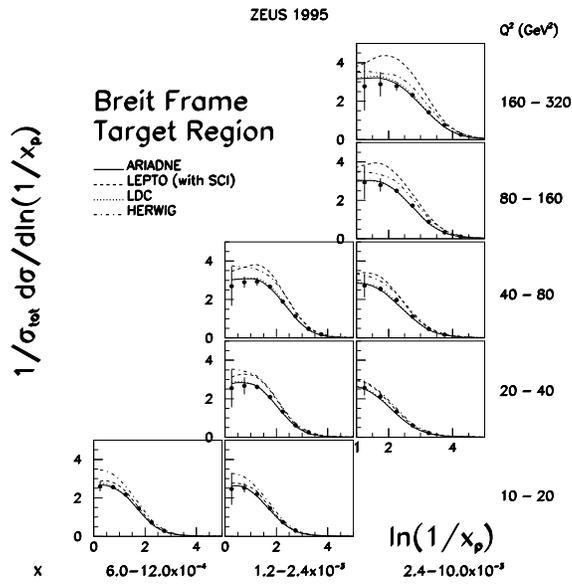,width=8cm}
\caption{Target fragmentation compared with models.}
\label{fig:Zeus_537_13}\end{center}\end{figure}
\end{itemize}

\section{Heavy quark fragmentation}\label{sec:heavy}

\begin{figure}\begin{center}\epsfig{file=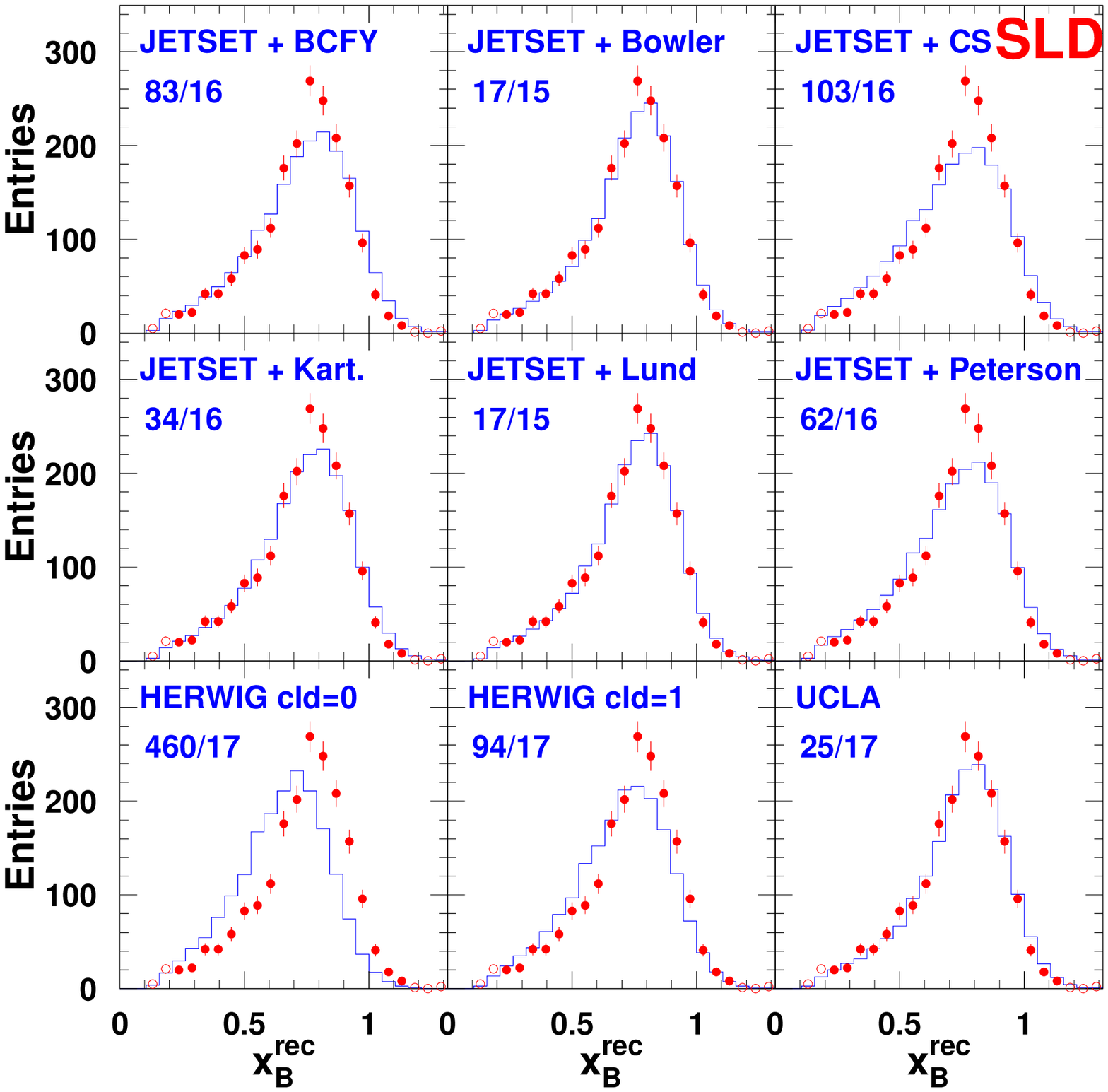,width=80mm}
\caption{SLD data on $b\to$ B fragmentation compared with models.}
\label{fig:SLD_8153_8}\end{center}\end{figure}

New data on $b\to$ B fragmentation from SLD \cite{Abe:1999fi},
using high-precision vertexing, discriminate between parton-shower
plus hadronization models (fig.~\ref{fig:SLD_8153_8}).
Note that the data have not yet been corrected for detector effects.

In general one expects the $b$ quark to lose energy by gluon emission
in the parton shower, and then to suffer a further non-perturbative energy
loss during hadronization. The latter is conventionally parametrized by
convoluting with the {\em Peterson function} \cite{Peterson:1983ak}:
\beq
f(z)= \frac 1z\left(1-\frac 1z-\frac{\eps_b}{1-z}\right)^{-2}
\;\;\;\;\;\;\;(z=x_B/x_b)\eeq
Including more perturbative QCD leads to a reduction in the amount of
non-perturbative smearing required to fit the data, and hence to a
smaller fitted value of the Peterson parameter $\eps_b$:

\noindent
Pure Peterson \cite{Abe:1999fi}: $\eps_b= 0.036$.\\
JETSET parton shower + Peterson \cite{Abe:1999fi}: $\eps_b= 0.006$.\\
NLLA QCD + Peterson \cite{Nason:1999zj}: $\eps_b= 0.002$.

\begin{figure}\begin{center}
\epsfig{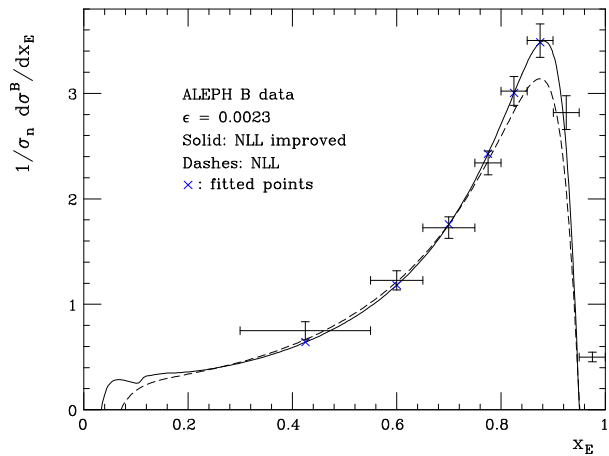}
\caption{ALEPH data on $b\to$ B fragmentation compared with NLLA QCD.}
\label{fig:aleph_nllimprov}
\end{center}\end{figure}

Here NLLA (next-to-leading logarithmic approximation) refers to
an analytical perturbative calculation that goes beyond the
parton shower approximation. The calculation of ref.~\cite{Nason:1999zj}
was fitted to earlier ALEPH data \cite{Buskulic:1995gp},
as shown in fig.~\ref{fig:aleph_nllimprov}.

In the universal low-scale $\as$ model, the perturbative prediction
is extrapolated smoothly to the non-perturbative region, with no Peterson
function at all \cite{Dokshitzer:1996ev}.

\section{Conclusions}\label{sec:conc}

We have seen that experimental studies of fragmentation are yielding
large amounts of new data for comparison with theoretical predictions
and models.  Especially impressive is the success of the MLLA perturbative
predictions in accounting for the general shape and energy-dependence of
fragmentation at small momentum fractions.  We are now at the stage when 
more detailed comparisons await new theoretical input, in the form of
coefficient functions that take account of selection procedures,
especially for gluon jets in $\ee$ final states.

Comparisons between data and hadronization models suggest that particle
masses, rather than quantum
numbers, are the dominant factor in suppressing heavy particle production.
Baryon production is not yet well described by any model.

Quark and gluon jets have the expected differences and these can be used
to measure the ratio of colour factors $C_A/C_F$. There is no strong evidence
yet for different particle content in gluon jets.

Fragmentation in deep inelastic lepton scattering shows some disagreements
with perturbative predictions. It is not yet clear whether these are due
to higher-order or non-perturbative effects.

New precise $b$ quark fragmentation data from Z$^0$ decay are now available
and put strong constraints on models for heavy quark hadronization. The
data suggest that perturbative effects dominate.

\section*{Acknowledgements}
It is a pleasure to congratulate the organisers for arranging such a
successful and truly international Summer School, and to thank the students
for making it so lively.

\end{document}